\documentclass[onecolumn,aps,prr,preprintnumbers,amsmath,amssymb]{revtex4-2}

\usepackage{natbib}
\usepackage{graphicx}
\usepackage{bm}
\usepackage{color}
\usepackage{float}
\usepackage{bbold}
\usepackage{amsmath}
\usepackage{amsfonts}
\usepackage{amssymb}
\usepackage{hyperref}
\usepackage{makeidx}
\usepackage{graphicx}

\newcommand{\ket}[1]{\left| #1 \right>}

\begin{document}
	
	\title{Non-Markovian amplitude damping in a central spin model with random couplings}

	\author{ Mehboob Rashid$^{1,2}$, Rayees A Mala$^2$, Saima Bashir$^1$, Muzaffar Qadir Lone$^2$\footnote{corresponding author: lone.muzaffar@uok.edu.in}}	

\affiliation{$^1$ Department of Physics, National Institute of Technology, Srinagar-190006 India,\\	$^2$ Quantum Dynamics Lab, Department of Physics, University of Kashmir, Srinagar-190006 India. }

\begin{abstract}
{	
	Non-Markovian dynamics is central to quantum information processing, as memory effects strongly influence coherence preservation, metrology, and communication. In this work, we investigate the role of stochastic system–bath couplings in shaping non-Markovian behavior of open quantum systems, using the central spin model within a time-convolutionless master equation framework. We show that the character of the reduced dynamics depends jointly on the intrinsic memory of the environment and on the structure of the system–environment interaction. In certain regimes, the dynamics simplify to pure dephasing, while in general both amplitude damping and dephasing contribute to the evolution. By employing two complementary measures: the Quantum Fisher Information (QFI) flow and the Breuer–Laine–Piilo (BLP) measure, we demonstrate that QFI flow may fail to witness memory effects in weak-coupling and near-resonant regimes, whereas the BLP measure still detects information backflow. Furthermore, external modulation of the interaction kernel produces qualitatively richer behavior, including irregular and frequency-dependent revivals of non-Markovianity. These results clarify the physical origin of memory effects, highlight the limitations of single-witness approaches, and suggest that stochasticity and modulation can be harnessed to engineer robust, noise-resilient quantum technologies.}
\end{abstract}

\maketitle
	\section{Introduction}
	
	The study of open quantum systems represents a crucial aspect of modern quantum physics, focusing on how quantum systems interact with their environments. These interactions are fundamental to advancements in quantum technologies, including quantum computing and quantum information processing \cite{1,2}. However, such systems face two primary challenges: \textit{decoherence} and \textit{dissipation}. Decoherence refers to the loss of quantum coherence due to interactions with the environment, degrading the system's quantum information. Dissipation, on the other hand, involves the transfer of energy from the system to the environment, which can irreversibly alter the system's dynamics. Addressing these challenges requires a deep understanding of both the Markovian and non-Markovian regimes, with the latter gaining attention for its incorporation of memory effects \cite{3,4}.  
	
	The concept of \textit{non-Markovianity} has been extensively studied, with various measures proposed to quantify it. Non-Markovianity arises from memory effects, where information lost to the environment flows back to the system, influencing its evolution \cite{5,6}. One widely used approach relies on the \textit{trace distance}, which tracks the distinguishability between two quantum states over time. In Markovian (memoryless) processes, the trace distance exhibits monotonic decay, whereas any non-monotonic behavior indicates information backflow, signaling non-Markovianity \cite{7,8,9}. Another important tool is the \textit{Quantum Fisher Information} (QFI), which measures the amount of information about a parameter extractable from system measurements. QFI serves as a witness to non-Markovian evolution by detecting instances of memory effects through information backflow. Together, these methods provide valuable insights into deviations from Markovian behavior \cite{10,11,12}.

	In this work, we demonstrate the emergence of non-Markovian amplitude damping in the central spin model with random system–bath couplings. The central spin model is a widely studied and versatile framework for exploring quantum decoherence, dissipation, entanglement, and general system-environment interactions \cite{14,15,16,17,18}. In this model, a central spin interacts with a surrounding ensemble of spins, making it directly applicable to physical systems such as electrons in quantum dots \cite{19,20,21,22,23,24,25,26,27} or nitrogen-vacancy centers in diamond \cite{nv1,nv2,nv3,nv4,nv5}. Beyond its relevance to specific systems, the central spin model plays a pivotal role in quantum information science, providing insights into qubit decoherence and dissipation while also guiding the design of quantum networks for computation and communication \cite{36,37}. By analyzing non-Markovian effects within this framework, we aim to elucidate how randomness in system-bath couplings and the structure of interactions can be exploited to control quantum dynamics and enhance memory effects in practical applications. {Our results show that the degree and character of non-Markovianity depend sensitively on  kernel structure of random couplings, with  modulated kernels producing richer, frequency-dependent revivals of information backflow than exponential kernels.}

	Advances in materials science, particularly in topological insulators and two-dimensional materials, have further expanded the relevance of this model. These materials often exhibit unpredictable and non-uniform couplings with the environment, arising from material-specific properties and random fluctuations \cite{38,39,40,41,42,43}. Recent investigations into non-Markovianity in central spin systems with random couplings have leveraged quantum simulation techniques, explored the emergence of persistent dark states, and utilized optimal control methods \cite{45,46,47}. Ultracold quantum simulators, for instance, enable precise control of interactions between a central spin and bath spins, allowing real-time observation of non-Markovian dynamics in highly controlled environments \cite{44,Gross,AP}. Furthermore, studies have shown that dark states can significantly enhance quantum memory by maintaining polarization over extended durations, even in disordered systems\cite{48,49}.  
	
	{The paper is organised as follows. In Sec. \ref{sec:Model}, we  introduce the central spin model with random couplings and its extension of reservoir fermion bath. Subsequently in  Sec. \ref{sec:dynamics} we derive the general Lindblad master equation for this system and solve it to obtain the general amplitude damping dynamics. In Sec. \ref{sec:MN} we consider  specific random coupling correlations and examine the non-Markovian nature of the dynamics for such models. We make our conclusions in Sec. \ref{sec:Concl}.}

	\section{Model Hamiltonian}\label{sec:Model}

	We consider a model in which a given qubit (system) with energey splitting $\omega_{0}$ is coupled to an ensemble of $N$ non-interacting spins (bath)  with each $k$-th spin having energy splitting $\omega_{k}$. {The system and bath togather are described by a $2^{N+1}$ dimensional Hilbert space $\mathcal{H} = \mathcal{H}_{s} \otimes \mathcal{H}_{b_1} \otimes \mathcal{H}_{b_2} \otimes \cdots \otimes \mathcal{H}_{b_N}$, where $\mathcal{H}_{s}$ and $\mathcal{H}_{b_k}$ pertain to the Hilbert space for the central spin and $k$-th spin bath, respectively \cite{2}.  The dynamics is governed by the Heisenberg  interaction with time dependent random couplings ($\hbar=1$):
	{	
	\begin{eqnarray}
				\mathbf{H} &=& \frac{\omega_0}{2}\, \sigma_0^z \otimes \mathbf{I}_{b} 
				+ \mathbf{I}_{2} \otimes \frac{1}{2} \sum_{k=1}^N \omega_k \, \sigma_k^z + \sum_{k=1}^N \frac{J_k(t)}{\sqrt{N}}\, 
				\vec{\sigma}_0 \cdot \vec{\sigma}_k 
			\end{eqnarray}
		where \( \vec{\sigma}_0 \cdot \vec{\sigma}_k := \sigma_0^x \sigma_k^x + \sigma_0^y \sigma_k^y + \sigma_0^z \sigma_k^z \) 
		denotes the standard scalar (dot) product between the Pauli operator vectors of the system and the \(k\)th bath spin, representing a Heisenberg-type exchange interaction.}
	
			 For simplicity, we will omit the explicit representation of identity operators $(\mathbf{I}_{b} , \mathbf{I}_{2})$ in the Hamiltonian from this point on and write 
	\begin{eqnarray}
		\mathbf{H}= \frac{\omega_0}{2}\sigma_0^z + \frac{1}{2}\sum_{k=1}^N \omega_k \sigma^z_k + \frac{1}{\sqrt{N}}\sum_{k=1}^N J_k(t)\vec{\sigma_0} \cdot \vec{\sigma}_k. \label{eq:H}
	\end{eqnarray}
	Now, we make use of the Jordan-Wigner transformation \cite{jw} (see App. \ref{apn:JWT}) on the spin bath in order to be able to use the  tools of quantum field theory to solve the system.  Under this transfomation, the spin operators  are mapped to the fermion operators $\hat{f}_{k}$ as 
	\begin{eqnarray}
		\sigma_k^z=2 \hat{f}^{\dagger}_k \hat{f}_k - \textbf{I}_{2}, ~~\sigma_k^+  = \sigma_{k}^{x} + i \sigma_{k}^{y} = \hat{f}_k^{\dagger},~~\sigma_k^- = \sigma_{k}^{x} -  i \sigma_{k}^{y} = \hat{f}_k, \label{eq:JWT}
	\end{eqnarray}
	where the fermion operators satisfy anti-commutation relation $\{f_k,f_{k^\prime}^{\dagger}\}=\delta_{k,k^{\prime}} \textbf{I}_{2}$. Under the mapping provided in Eq.~(\ref{eq:JWT}), the Hamiltonian in Eq.~(\ref{eq:H}) can be expressed as a sum of the system Hamiltonian ($\mathbf{H}_0$), the bath Hamiltonian ($\mathbf{H}_B$), and the interaction Hamiltonian ($\mathbf{H}_I$), such that $	\mathbf{H}' = \mathbf{H}_{S} + \mathbf{H}_B + \mathbf{H}_I$ where
	\begin{align}
		\mathbf{H}_{S} &=  \frac{\omega_r}{2}\sigma_0^z, \quad   \mathbf{H}_{B} =\sum_{k=1}^N \omega_k \hat{f}^{\dagger}_k \hat{f}_k,  \label{eq:HSHB}\\
		\mathbf{H}_{I} &=   \frac{1}{\sqrt{N}}\sum_k J_k(t) \left[ \sigma_0^+  \hat{f}_k + \sigma_0^-   \hat{f}^{\dagger}_k + 2\sigma_0^z  \hat{f}^{\dagger}_k \hat{f}_k \right] \label{eq:Hi}
	\end{align}
	Note that the energy of the system qubit is shifted due to interaction with the spin bath and is given by  $\omega_r= \omega_0 - \frac{2}{\sqrt{N}}\sum_k J_k(t)$. {It is also  evident that the interaction Hamiltonian does not commute with the system Hamiltonian, which inherently suggests the occurrence of  \emph{dissipation} in the system.}

	\section{Dynamics}\label{sec:dynamics}
	
	Let us choose a spin basis $\{ |\uparrow \rangle, | \downarrow \rangle  \}$  and write the initial state of the system as $|\psi_S(0)\rangle= a|\uparrow\rangle+ b|\downarrow\rangle$ where $a,b \in \mathbb{C}$ with $|a|^2 + |b|^2 =  1$. The bath spins are all assumed to be in $\ket{\downarrow}$ state so that the bath state is written as  
		$|\psi_B\rangle =  \prod_{k=1}^N |\downarrow\rangle_k$.  Assuming the system and bath are uncorrelated initially, we write the total state as the product of the two states
		\begin{equation}
			|\Psi_T(0)\rangle= |\phi_S(0)\rangle \otimes |\psi_B\rangle = \left( a|\uparrow\rangle+ b|\downarrow\rangle \right) \otimes \prod_{k=1}^N |\downarrow\rangle_k.
		\end{equation}
{		As mentioned in the previous section, we are interested in mapping the spin bath to a fermion bath. The bath spin state written above translates into the fermion vacuum state:   $|\psi_B\rangle = \prod_{k=1}^N |\downarrow\rangle_k = \prod_{k=1}^N |0\rangle_k \equiv  |0_B\rangle$.  It is often desirable to select the bath in its vacuum state to simplify the averaging process over the bath state. 
	}

	Having set the initial state of the total system, we make use of the well know master equation under Born approximation\cite{2}
	\begin{equation}
		\label{eq:ME}
		\frac{\partial }{\partial t} \rho^S_{int}(t) =
		- \int_0^t dt' {\rm Tr_B}[\mathbf{H}_{int}(t),[\mathbf{H}_{int}(t'), \rho^S_{int}(t) \otimes |0_B\rangle \langle 0_B|]],
	\end{equation}
	where $\mathbf{H}_{int}(t)$ is the interaction Hamiltonian $	\mathbf{H}_{I} $ in Eq.~(\ref{eq:Hi}) written in the interaction picture and is given by (see App. \ref{apn:Hint} for details)
\begin{align}
	\mathbf{H}_{int}(t) = \frac{1}{\sqrt{N}} \sum_k J_k(t) \big[ &
	\sigma^+ f_k e^{i (\omega_0 - \omega_k)t} + \sigma^- f^{\dagger}_k e^{-i (\omega_0 - \omega_k)t} + 2\sigma^z f^{\dagger}_k f_k 
	\big].
	\label{eq:Hint}
\end{align}
Further  simplification transforms Eq.~(\ref{eq:ME})  into the Lindblad form  (as derived in App. \ref{apn:LindbladEqn})
\begin{align}
	\frac{\partial }{\partial t} \rho^S(t) = 
	& -i [\mathbf{h}_S(t), \rho^S(t)] + \gamma(t) \big[ 
	2 \sigma^{-} \rho^S(t) \sigma^{+} 
	- \sigma^{+} \sigma^{-} \rho^S(t) 
	- \rho^S(t) \sigma^{+} \sigma^{-} 
	\big].
	\label{MAS1}
\end{align}
where the time dependent system Hamiltonian $\mathbf{h}_S(t)$ and the decay rate $\gamma(t)$ that depends on the couplings $J_k(t)$ as follows
\begin{align}
	\mathbf{h}_S(t) &= \omega_r(t) \sigma^z  - \left[ \int_0^{t} dt' \sum_k J_k(t) J_k(t') 
	\sin\Delta_k(t - t') \right] \sigma^- \sigma^+ \notag \\
	\gamma(t) &= \frac{1}{N} \int_0^{t} dt' \sum_k J_k(t) J_k(t') 
	\cos\Delta_k(t - t').
\end{align}

	We can also define the function $\gamma(t)$ through a total correlation function, $\alpha(t) = \gamma(t) + i \phi(t)$, where
	\begin{eqnarray}
		\phi(t) =   \frac{1}{N} \int_0^{t}d t' \sum_k J_k(t)J_k(t')\sin\Delta_k(t-t')
	\end{eqnarray}
	We observe that the time evolution of  any  system density matrix governed by the above equation \ref{eq:ME} depends on the random couplings $J_k(t)$. Therefore, we need to average over these couplings with specific distributions. Therefore, averaging over these random variables, the  time evolution of an initial  qubit density matrix $\rho_S(0)=|\phi_S(0)\rangle \langle \phi_S(0)|$, can be written as follows (see App. (\ref{apn:Sol}) for the derivation)
	\begin{eqnarray}\label{dm}
		\rho^S(t) =\begin{pmatrix}
			|a|^2 F_0(t) & ab^{\star}F_1(t)\\
			a^{\star}b F_1^{\star}(t)&1-|a|^2 F_0(t)
		\end{pmatrix},
	\end{eqnarray}
	where $F_1(t)= e^{-[\Gamma(t)+ i \Phi(t)]}$, $ F_0(t)=|F_1(t)|^2= e^{-2\Gamma(t)}$. Here $\Gamma(t) = \Re \int_0^{t} d z \langle \alpha(z)\rangle $ and $\Phi(t) = \Im \int_0^{t} d z \langle \alpha(z)\rangle $ where
	
	\begin{eqnarray}
		\alpha(z)= \frac{1}{N}\sum_k \int_0^{z} d t'  ~\overline{ J_k(z) J_k(t') } e^{i (\omega_0-\omega_k)(z-t')},
	\end{eqnarray}
	$\overline{ J_k(z) J_k(t')} $ represents the average over given distribution of  random couplings.
	{
	The dynamics presented by the equations \ref{MAS1}and \ref{dm} can be represented by the amplitude damping channel with Kraus operators
	\begin{eqnarray}
	K_0= \begin{pmatrix}
		F_1(t) & 0\\
		0& 1
	\end{pmatrix},
	~~K_1= \begin{pmatrix}
		0 & 0	 \\
\sqrt{1-F_0(t)}& 0
	\end{pmatrix}
\end{eqnarray} 
	such that $\rho_S(t)= \sum_{m=0}^1K_m\rho_S(0)K_m^{\dagger}$.}
	Next, in order to simplify the calculations, we assume the bath spins have the same energy $\omega_k=\omega$ for all $k$.  
	We now consider different ways $J_k(z)$'s are correlated due to their randomness  to calculate $\alpha(z)$. First, as a consistency check, we assume  Markovian noise i.e when the couplings are delta correlated:  $\langle J_k(z) J_k(t') \rangle= \kappa \delta(z-t') $, with decay rate $\kappa$. In this case, we  have
	\begin{eqnarray*}
	\alpha(z)= \frac{1}{N}\sum_k \int_0^z  d t' \kappa \delta(z- t') e^{i(\omega_0-\omega)(z-t')} = \frac{\kappa}{N}.
	\end{eqnarray*} 
		{
	This results in $\rho_{\uparrow\uparrow}(t)= \rho_{\uparrow \uparrow}(0)e^{-2\kappa t} = \rho_{\uparrow \uparrow}(0)e^{-t/t'_1}$ and $ \rho_{\downarrow\uparrow}(t)=\rho_{\downarrow\uparrow}(0) e^{-\kappa t}=\rho_{\downarrow\uparrow}(0) e^{- t/t'_2} $ with $t'_2=2t'_1= 1/\kappa$ in conformity with the Markovian result.

		In this work, we investigate two classes of noise correlations that model environmental memory effects in open quantum systems: the exponential memory kernel and the modulated (oscillatory) memory kernel. These models determine the time-correlated behavior of the bath couplings $\overline{J_k(t) J_k(t')}$, and play a crucial role in shaping the dynamics of decoherence, entanglement, and non-Markovianity\cite{69}.		
		The noise correlation function is given by
			\begin{eqnarray}
			\overline{ J_k(t) J_k(t') }=  \begin{cases} 
				\kappa e^{-\kappa(t-t')}~~~\text{Exponential Memory Kernel}, \\
				\kappa e^{-\kappa(t-t')} \cos \omega(t-t') ~~\text{Modulated Kernel}.
			\end{cases}
		\end{eqnarray}
		where $\kappa$ sets the memory decay rate and $\omega$ introduces oscillations that can lead to coherence revivals. 		
		The exponential kernel corresponds to environments with memory that decays exponentially in time, typical of systems with Lorentzian spectral densities. It characterizes a semi-Markovian regime  where correlations decay monotonically.
		The modulated kernel introduces an oscillatory component to the memory function, representing environments with structured spectral densities or feedback effects . This model has been used to capture more complex memory structures, such as those arising in spin-boson or photonic environments.		
		These two models allow us to explore a range of physical scenarios from relatively weak memory effects to strong non-Markovian dynamics with revivals and their impact on coherence and quantum information protocols.
		
	}

	Before proceeding further, we introduce the key parameters and notations used in our analysis. The system is characterized by the detuning $ \Delta = \omega_0 - \omega $, the decay rate $ \kappa $, and the modulated energy $ \omega $. To facilitate calculations, we first express $ \gamma(t) $ and $ \Phi(t) $ in terms of $ \Delta $, $ \kappa $, and $ \omega $, and then redefine them using the dimensionless parameters  	
	\begin{equation}
		\bar{\Delta} = \frac{\Delta}{\omega}, \quad \Bbbk = \frac{\kappa}{\omega}, \quad \tau = \omega t.
	\end{equation}	
	This reformulation provides a more convenient scaling for the problem. Further details on this transformation are provided in Appendix \ref{apn:GammaPhi}.
		 We now plot the population difference { $\mathcal{P}_D=\rho_{\uparrow\uparrow}-\rho_{\downarrow \downarrow}$ }and coherence $\mathcal{C}$  for different parameter regimes  with  initial state $|\phi_S(0)\rangle=\frac{1}{\sqrt{2}}[|\uparrow\rangle+ |\downarrow\rangle]$. The coherence is measured using $l_1$-norm defined as $\mathcal{C}=\sum_{i\ne j}|\rho_{ij}| $, which represents the sum of  magnitude of the off-diagonal elements  of the density matrix. In figure \ref{figurea}, we have plotted coherence for different $\Bbbk$, $\bar{\Delta}$ values. In fig. \ref{figurea}(a) and \ref{figurea}(b), we have coherence $\mathcal{C}$ plotted for exponential memory kernel while \ref{figurea}(c) and \ref{figurea}(d) represent the variation of $\mathcal{C}$ for modulated kernel. Also, we plotted the population difference $\mathcal{P}_D$ for these two memory kernels in figure \ref{figureb}(a)-(d). We observe from these graphs as the $\mathcal{P}_D$  saturates to 1, correspondingly the $\mathcal{C}$ decays to zero. For example, in case of exponential memory kernel, for $\bar{\Delta}=0$ depicted in figures \ref{figurea}(a) and \ref{figureb}(a),  we see that the coherence vanishes for all the values of $\Bbbk$ while the $\mathcal{P}_D$ saturates to a value of 1. This phenomena can be attributed to the fact as the system evolves from the initial state $|\phi_S(0)\rangle$, due to decoherence, the system gets localized in one of the eigen states and thus causing zero coherence in the system. Same trend is followed by all other graphs. Furthermore, we observe that for smaller values of $\Bbbk$ yield longer coherence in comparison to its larger values. Thus tuning parameter  $\Bbbk$ can lead to non-Markovian behaviour in the qubit dynamics. In case of modulated kernel, we can tune the system for longer coherences. In figures \ref{figurea}(c)-(d) and \ref{figureb}(c)-(d), we see that for the values of $\Bbbk=0.01$, the system maintains coherence over long times reflecting the feature of non-Markovianity in the time evolution.  This hold true, if we calculate the non-Markovianity of the underlying dynamics as shown in next section.

	\begin{figure}[t]
		\includegraphics[width=4.25cm,height=4cm]{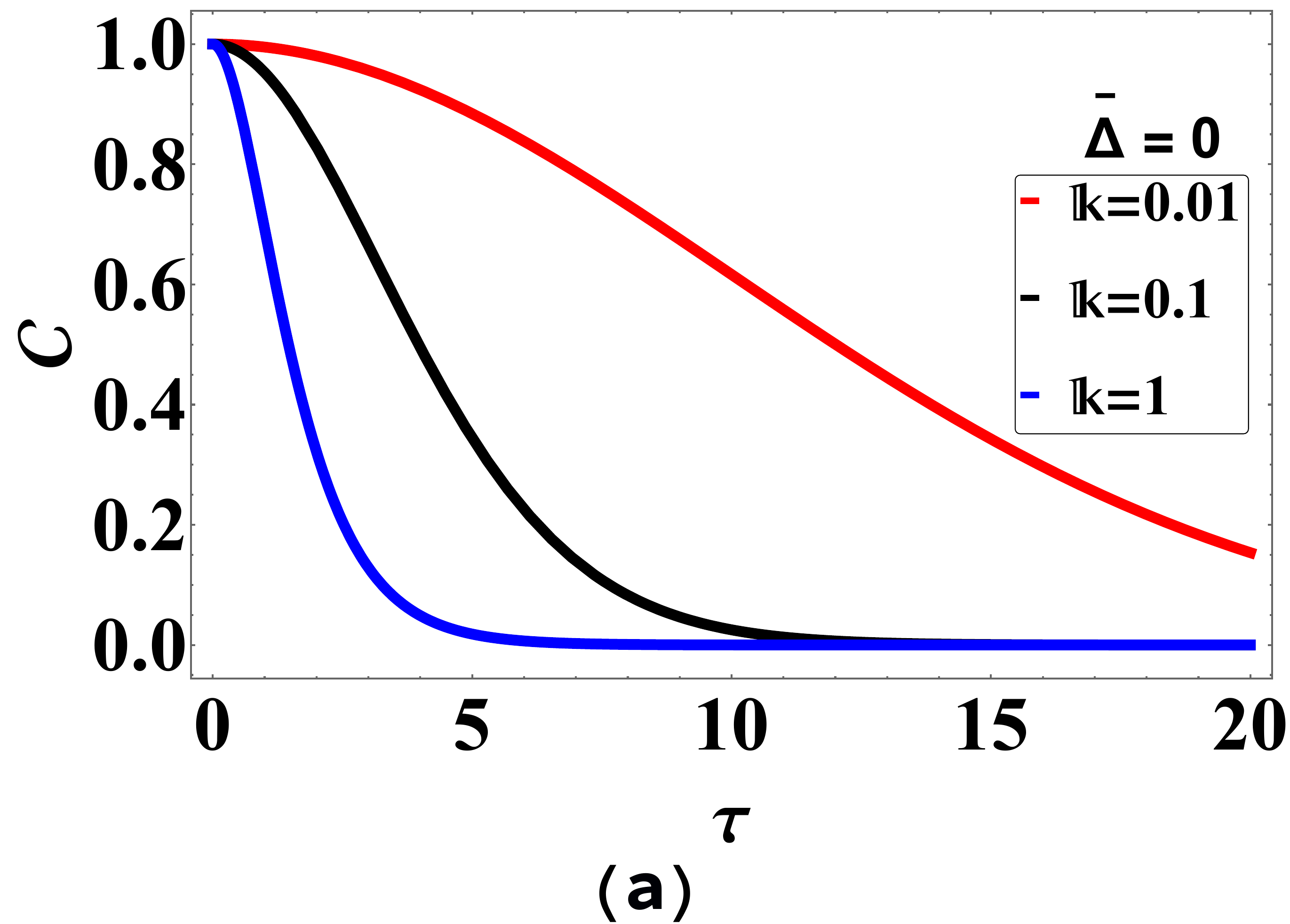}
		\includegraphics[width=4.25cm,height=4cm]{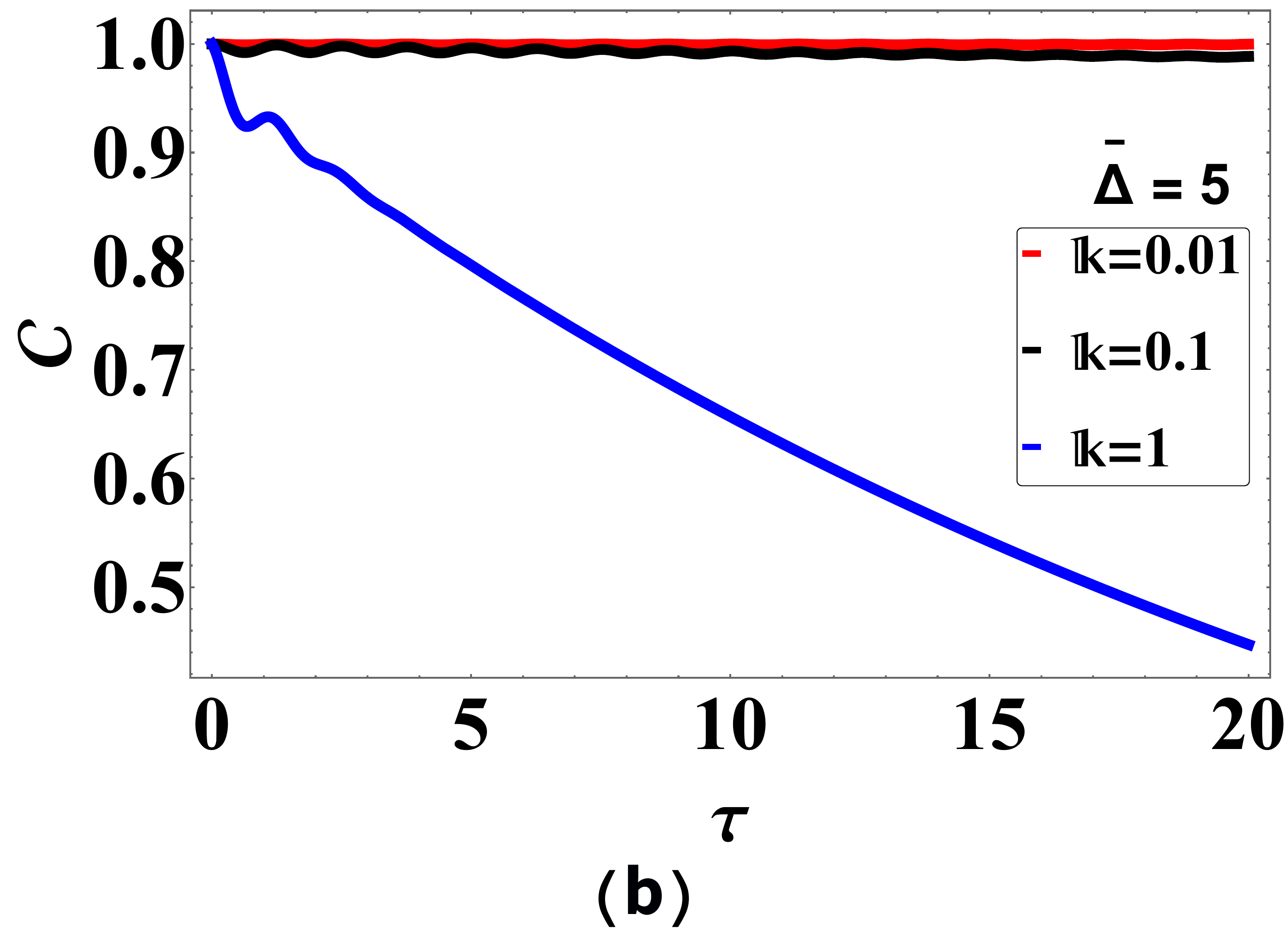}
		\includegraphics[width=4.25cm,height=4cm]{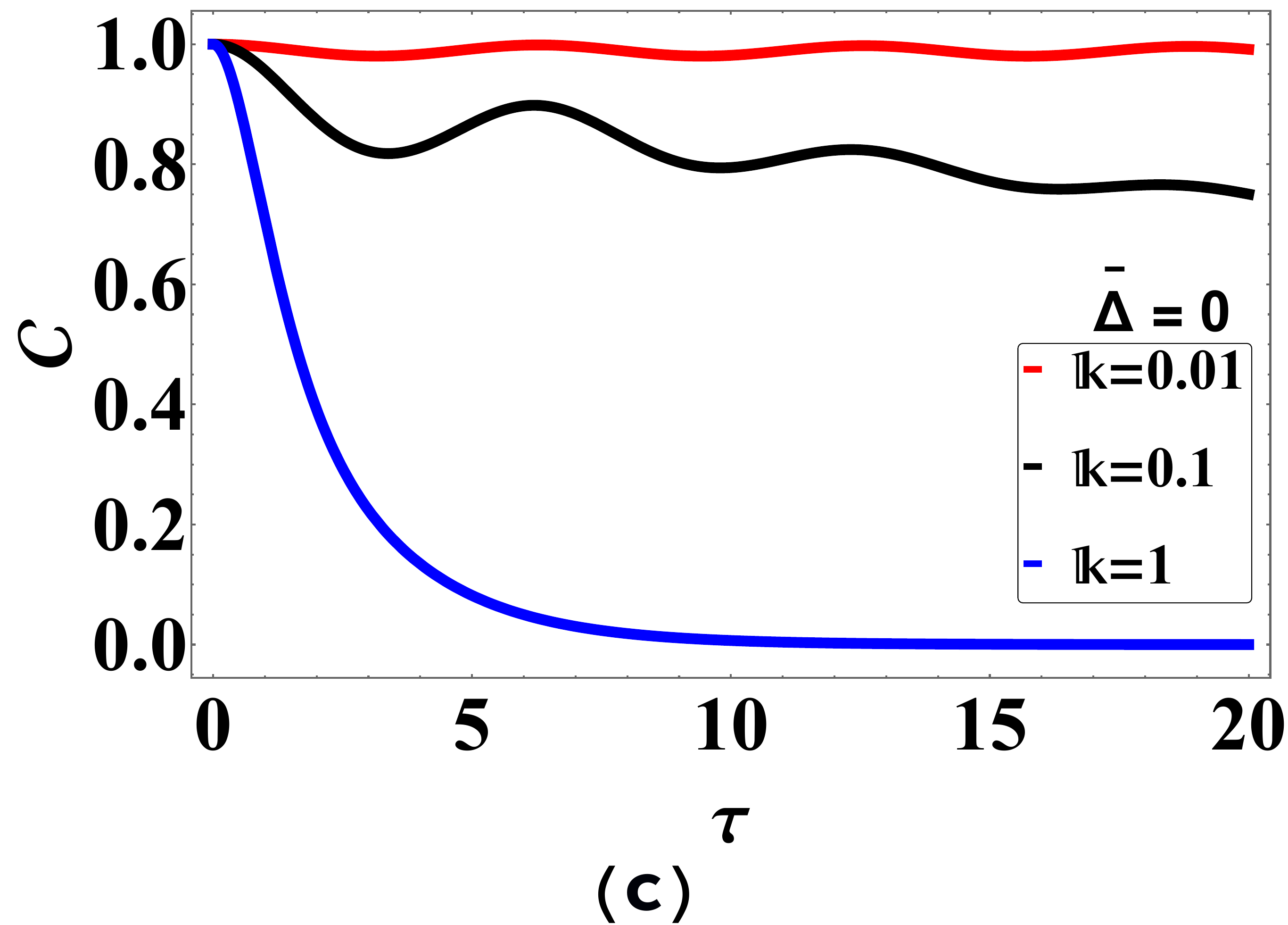}
		\includegraphics[width=4.25cm,height=4cm]{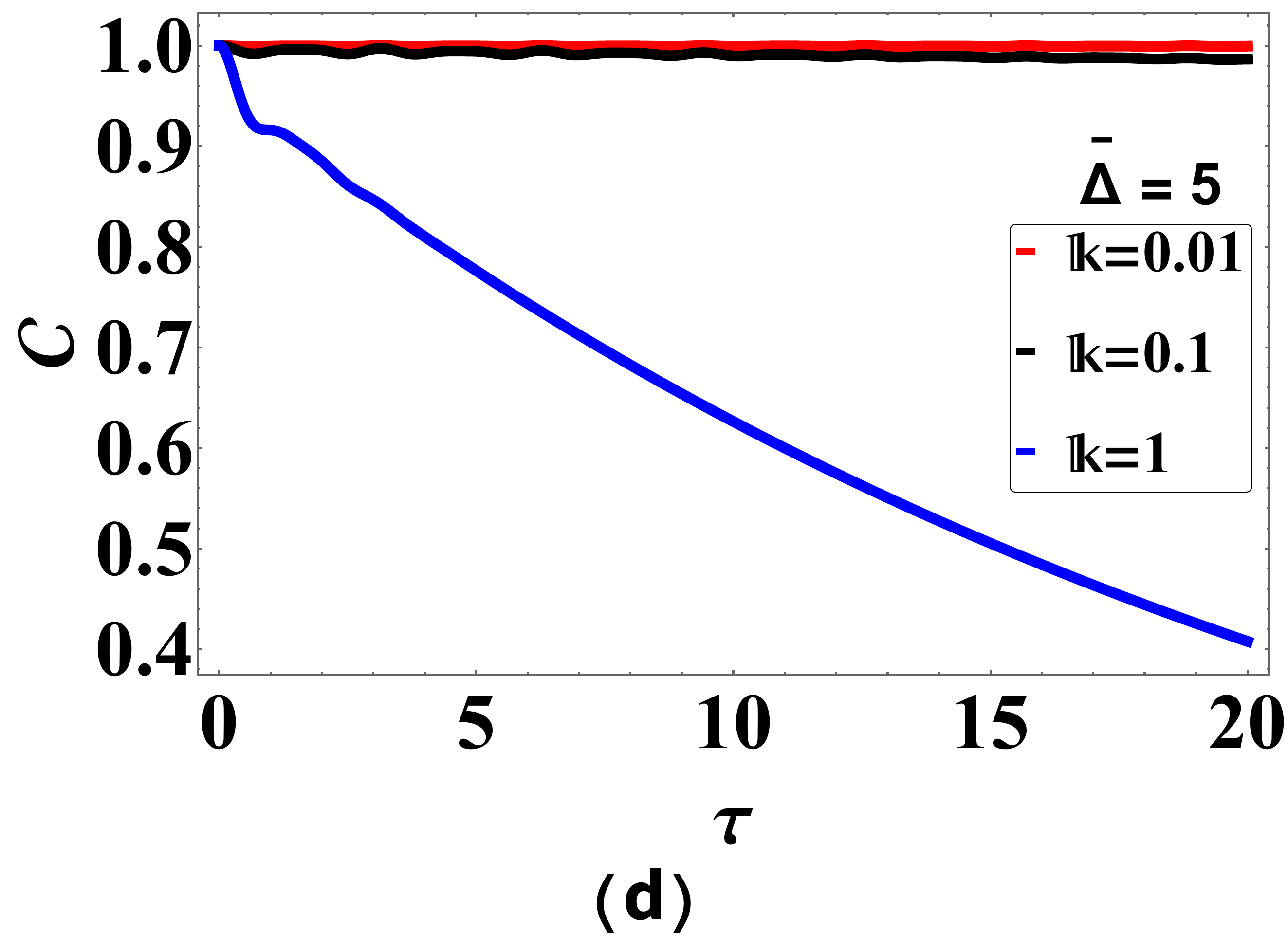}		
		\caption{Variation of coherence with $\tau$ for different parameter values. (a) and (b) represent coherence for exponential memory kernel while (c) and (d) are for modulated kernel. The parameter $\Bbbk$ is chosen in way such that the  dynamics has non-Markovian behavior. In each case, $\bar{\Delta}=0, 5 $  implying resonance and off-resonance cases respectively. }
		\label{figurea}
	\end{figure}
	\begin{figure}[t]
	\includegraphics[width=4.25cm,height=4cm]{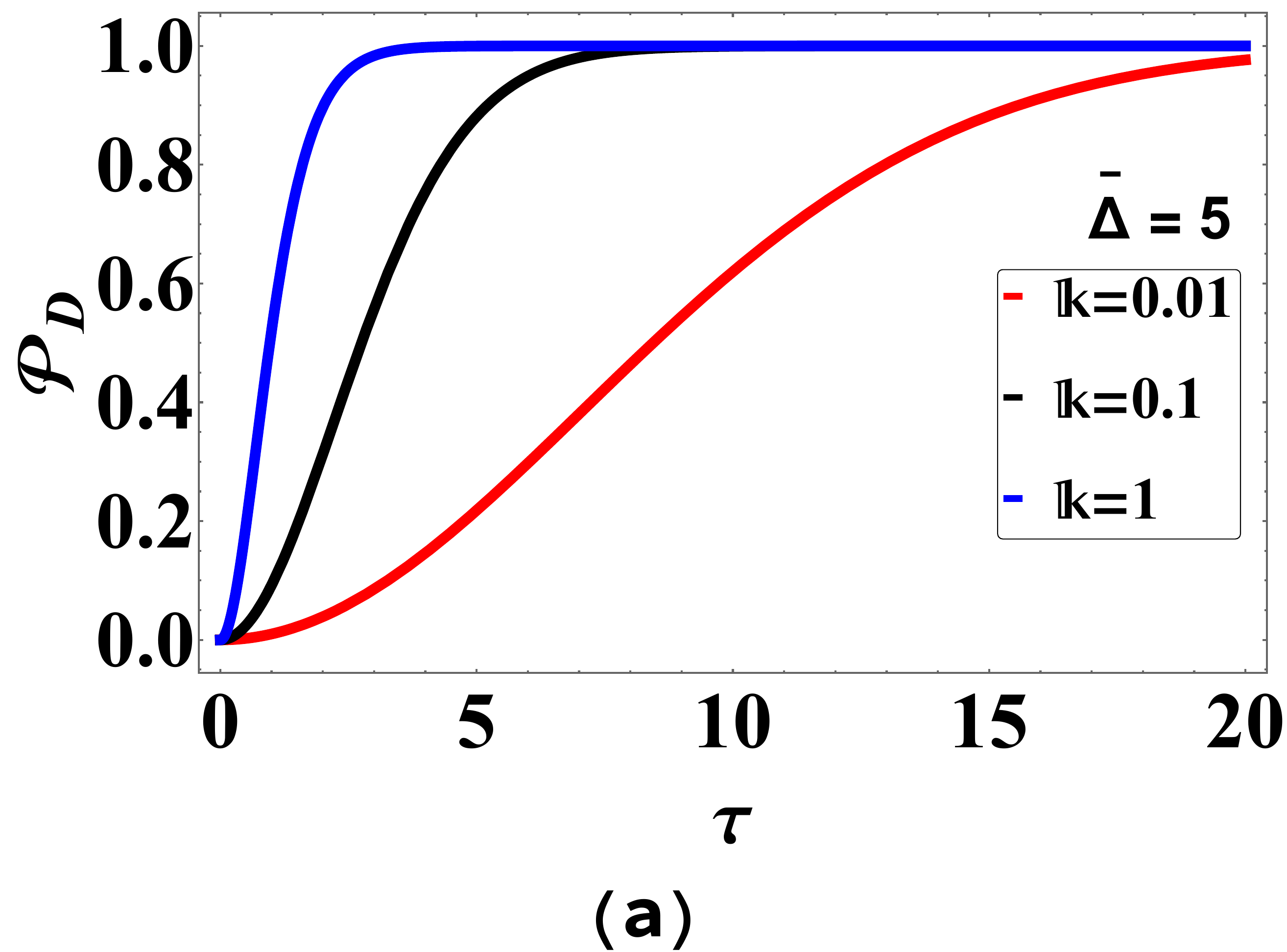}
		\includegraphics[width=4.25cm,height=4cm]{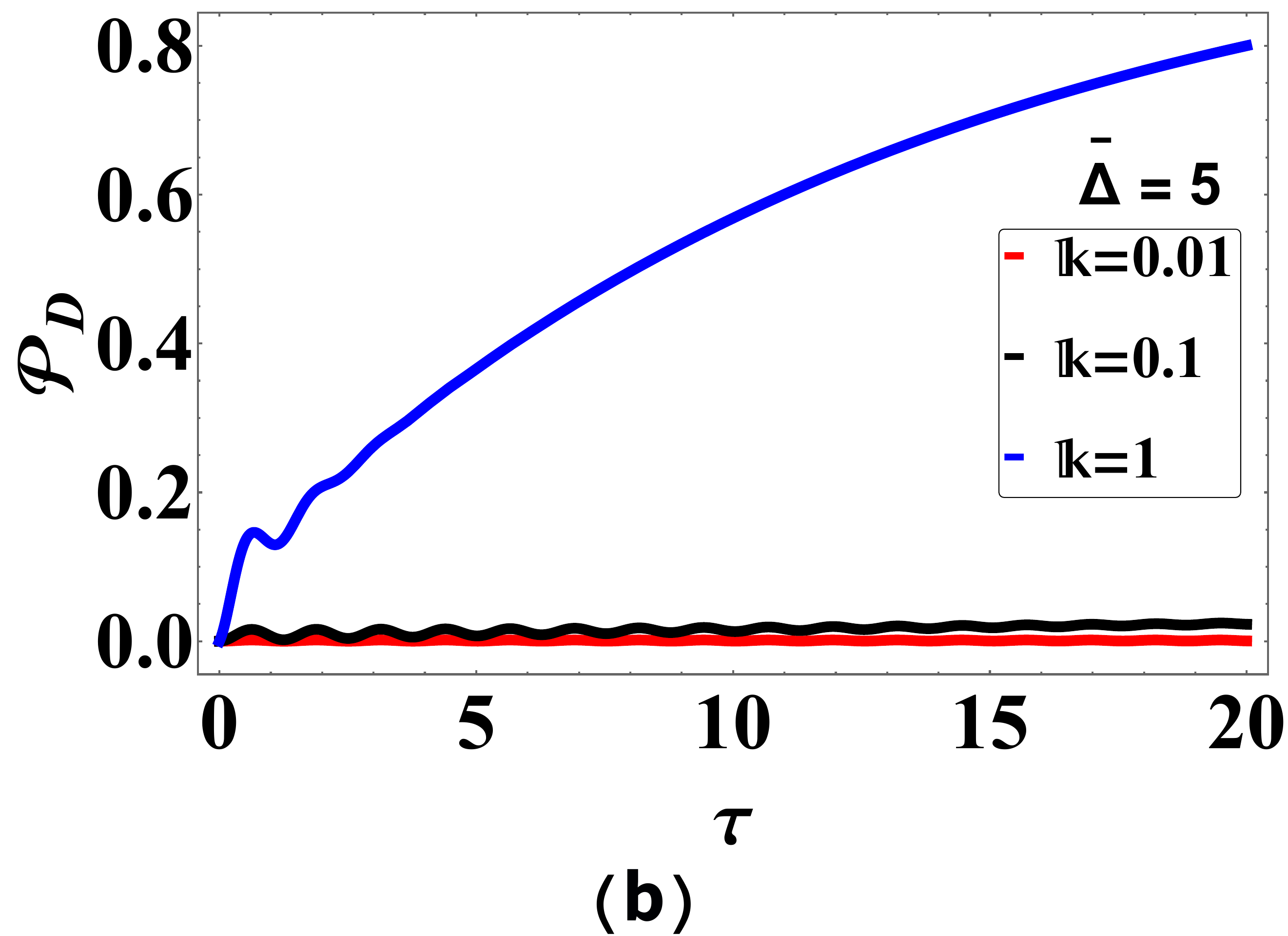}
		\includegraphics[width=4.25cm,height=4cm]{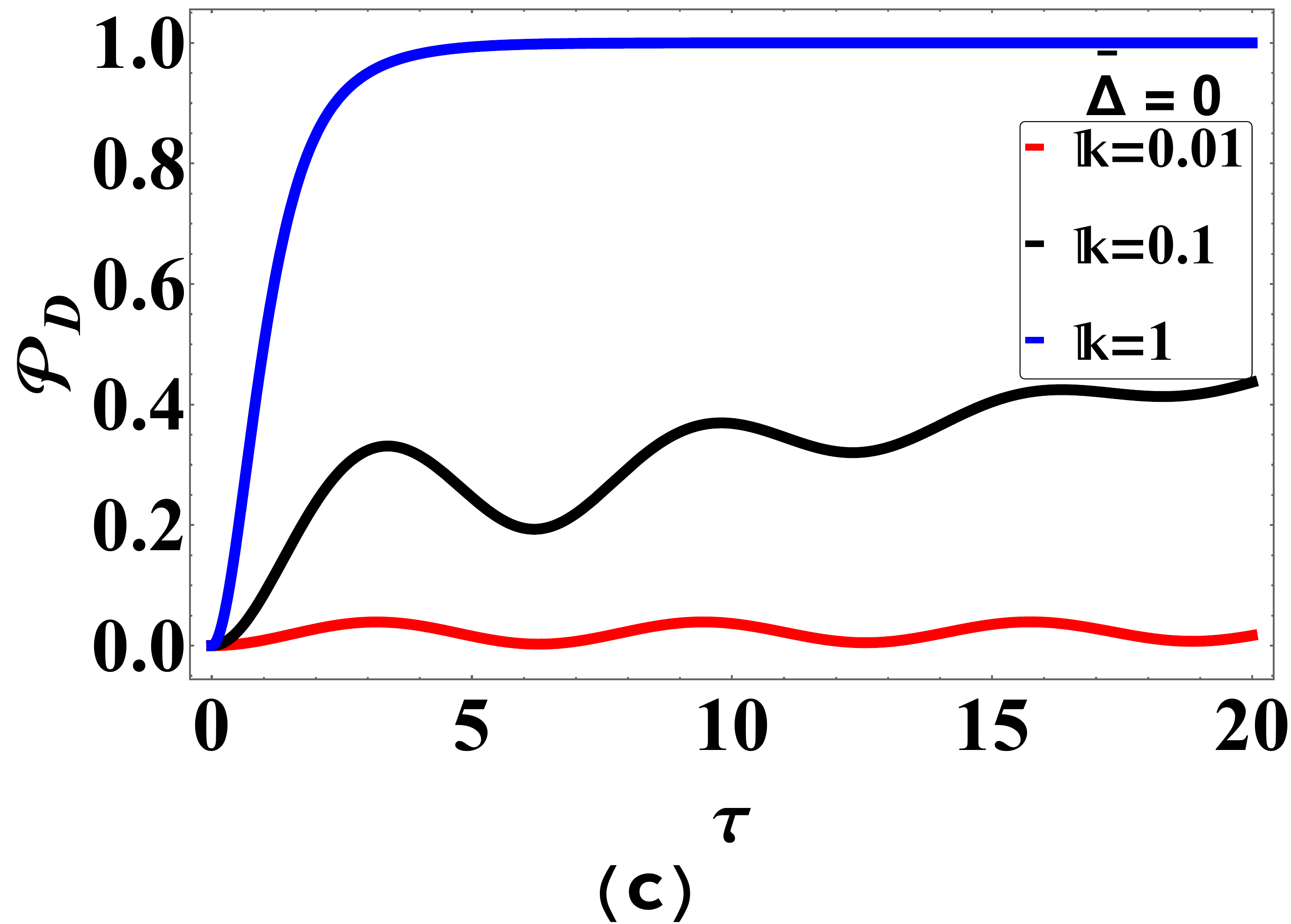}
		\includegraphics[width=4.25cm,height=4cm]{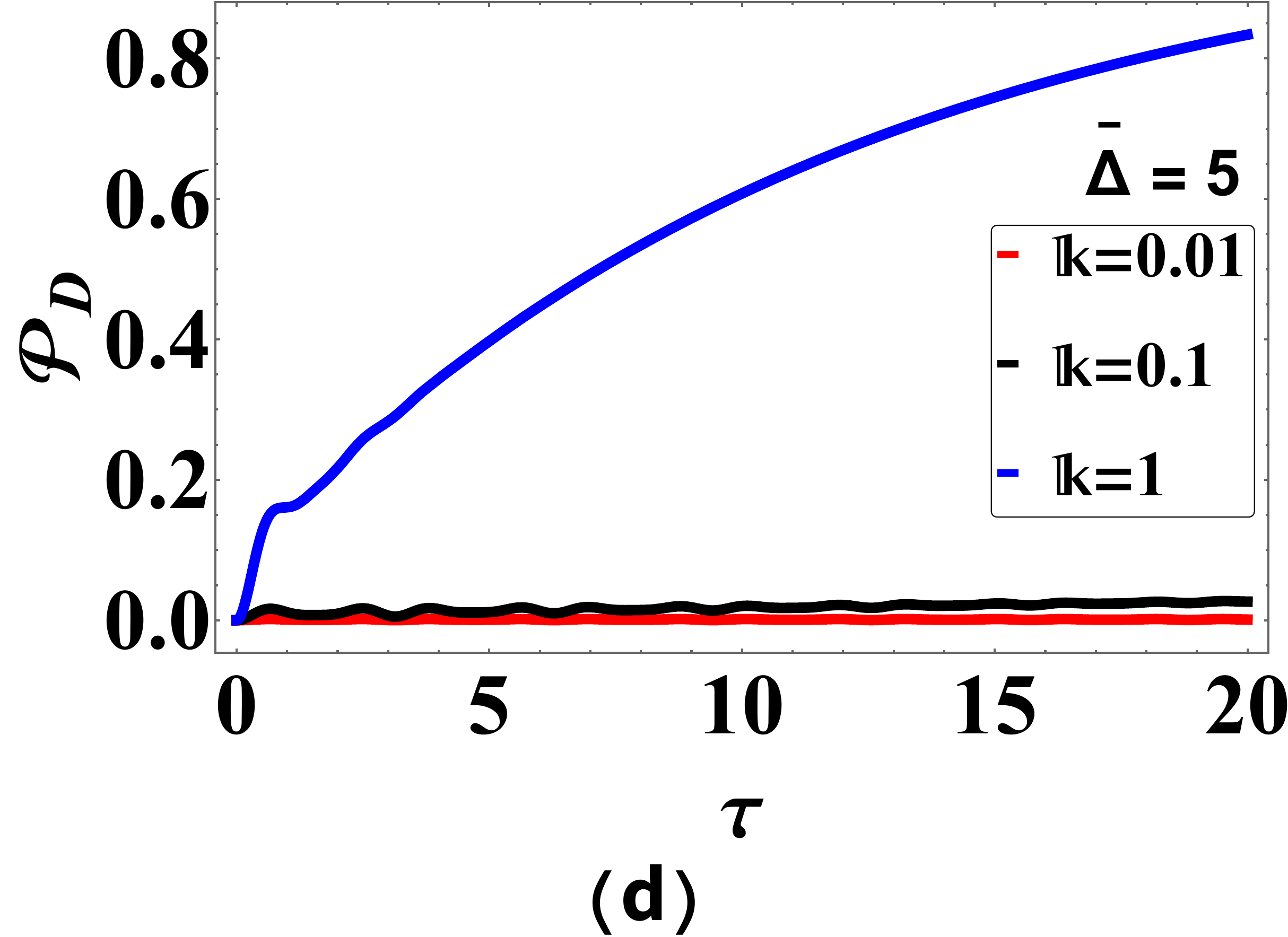}		
		\caption{Variation of population difference $\mathcal{P}_D$ with $\tau$ for different parameter values. (a) and (b) represent $\mathcal{P}_D$ for exponential memory kernel while (c) and (d) are for modulated kernel. The parameter $\Bbbk$ is chosen in way such that the  dynamics has non-Markovian behavior. In each case, $\bar{\Delta}=0, 5 $  implying resonance and off-resonance cases respectively. }
		\label{figureb}
	\end{figure}

	\section{Probing non-Markovian dynamics}\label{sec:MN}
	
	\subsection{Quantum Fisher Information and Its Role in Detecting Non-Markovianity}  
	
	Quantum Fisher Information (QFI) is a cornerstone of quantum metrology, quantifying the precision in estimating an unknown parameter $\theta$  encoded in a  quantum state $\rho_{\mu}$  via quantum Cramer-Rao inequality $\Delta^2 \mu \ge 1/n F(\rho_{\mu})$, where $\Delta^2 \mu$ is the variance, $F(\rho_{\mu})$ the QFI and $n$ is the number of  measurements. Thus, QFI reflects the ultimate precision achievable in estimating a parameter through quantum measurements \cite{47,48,49}. 
	For the single qubit density matrix parameterized by $\mu$, and Bloch vector $\vec{r}(\mu) $, QFI can be written in terms of the formula given below \cite{53}:
	\begin{eqnarray}\label{qfi}
		F(\rho(\mu), A) = \frac{\left[\vec{r}(\mu) \cdot \partial_\mu \vec{r}(\mu)\right]^2}{1 - \left|\vec{r}(\mu)\right|^2} + \left|\partial_\mu \vec{r}(\mu)\right|^2,
	\end{eqnarray}
where	$\partial_\mu \vec{r}(\mu) $ represents its derivative with respect to the parameter $\mu$. Beyond its applications in metrology, QFI serves as a powerful tool for analyzing the dynamics of quantum systems, offering insights into their evolution and sensitivity to environmental interactions.  
	
	In the study of open quantum systems, QFI becomes particularly valuable for detecting memory effects associated with non-Markovian dynamics. The concept of \textit{QFI flow} $\mathcal{F}_{\mu} = d F(\rho_{\mu})/dt$, which describes the rate of change of QFI over time, provides a dynamic measure of information transfer between the system and its environment. In Markovian processes, QFI typically decreases monotonically due to irreversible information loss. However, in non-Markovian dynamics, a positive QFI flow indicates the recovery of information previously lost to the environment, signifying memory effects \cite{9,55}:
	\begin{eqnarray}
		\mathcal{F}_\mu  =  \begin{cases} 
			\leq 0 & \text{Markovian dynamics}, \\
			> 0 & \text{Non-Markovian dynamics}.
		\end{cases}
	\end{eqnarray}
	This non-monotonic behavior in QFI flow acts as a clear signature of non-Markovianity, offering a robust framework for characterizing the influence of the environment on system evolution.

	We now analyze the non-Markovianity of the single qubit dynamics  as characterized by the flow of quantum Fisher information.  Consider an arbitrary qubit state at the initial time $t=0$, described by the state:~~{$|\psi\rangle =\cos\frac{\theta}{2}|\uparrow\rangle + \sin \frac{\theta}{2} e^{i \nu} |\downarrow\rangle$.}
	
		The initial Bloch vector of this qubit is represented as $\vec{r}(0)=(\frac{1}{2}\cos\nu\sin\theta, \frac{1}{2}\sin\nu\cos\theta, \frac{1}{2}\cos\theta)$. Under the specified central spin model, the density matrix evolves in accordance with equation (\ref{dm}). Consequently, the time-dependent Bloch vector is expressed as $\vec{r}(t)=(r_x(t),r_y(t),r_z(t))$ where
		\begin{eqnarray}
			r_x(t) = \frac{1}{2} \sin\theta\cos(\nu+ \Phi(t))e^{-\Gamma(t)}, ~~
			r_y(t) = \frac{1}{2}\sin\theta \sin(\nu+ \Phi(t))e^{-\Gamma(t)},~~
			r_z(t) = \frac{1}{2}\cos\theta e^{-2\Gamma(t)}.
		\end{eqnarray}
		Next, we see that
		\begin{eqnarray}
			\vec{r}(t). \frac{\partial \vec{r}(t) }{\partial \nu}= 0 ~~ {\rm and }~~ \vec{r}(t). \frac{\partial \vec{r}(t) }{\partial \theta} = \frac{1}{4}\sin\theta\cos\theta e^{-2\Gamma(t)}\big(1-e^{-2\Gamma(t)}\big) .
		\end{eqnarray}
		Consequently, applying Equation (\ref{qfi}), we can derive the Quantum Fisher Information (QFI) for the parameters $\theta$ and $\phi$ as :
	\begin{align*}
		F_\theta &= \frac{\left[\frac{1}{4} \sin\theta \cos\theta \, e^{-2\Gamma(t)} \left(1 - e^{-2\Gamma(t)}\right)\right]^2}
		{1 - \left(\frac{1}{4} e^{-2\Gamma(t)} \left( \sin^2\theta + \cos^2\theta \, e^{-2\Gamma(t)} \right)\right)} 
		+ \frac{1}{4} e^{-2\Gamma(t)} \left( \cos^2\theta + \sin^2\theta \, e^{-2\Gamma(t)} \right) \\
		F_\nu &= \frac{1}{4} e^{-2\Gamma(t)} \sin^2\theta
	\end{align*}
	and the associated quantum Fisher Information (QFI) flow is expressed as:
	\begin{equation}
		\boxed{
			\begin{aligned}
				\mathcal{F}_\theta &= 
				-\frac{
					e^{-2 \Gamma(t)} 
					\Big( 
					1 + 2 e^{2 \Gamma(t)} - 24 e^{4 \Gamma(t)} + 32 e^{6 \Gamma(t)} 
					+ 16 e^{8 \Gamma(t)} 
					+ \big( 
					1 - 2 e^{2 \Gamma(t)} + 8 e^{4 \Gamma(t)} 
					- 32 e^{6 \Gamma(t)} + 16 e^{8 \Gamma(t)} 
					\big) \cos(2 \theta) 
					\Big) 
					\frac{d\Gamma(t)}{dt}
				}
				{4 \big( 
					-4 e^{4 \Gamma(t)} + \cos^2(\theta) + e^{2 \Gamma(t)} \sin^2(\theta) 
					\big)^2} \\[10pt]
				\mathcal{F}_\nu &= 
				\frac{d F_\nu}{dt} = 
				-\frac{1}{2} \sin^2\theta \, e^{-2 \Gamma(t)} \frac{d\Gamma(t)}{dt}
			\end{aligned}
		}
		\label{eq:boxed_eqn}
	\end{equation}

	\begin{figure}[t]
		\includegraphics[width=4.25cm,height=4cm]{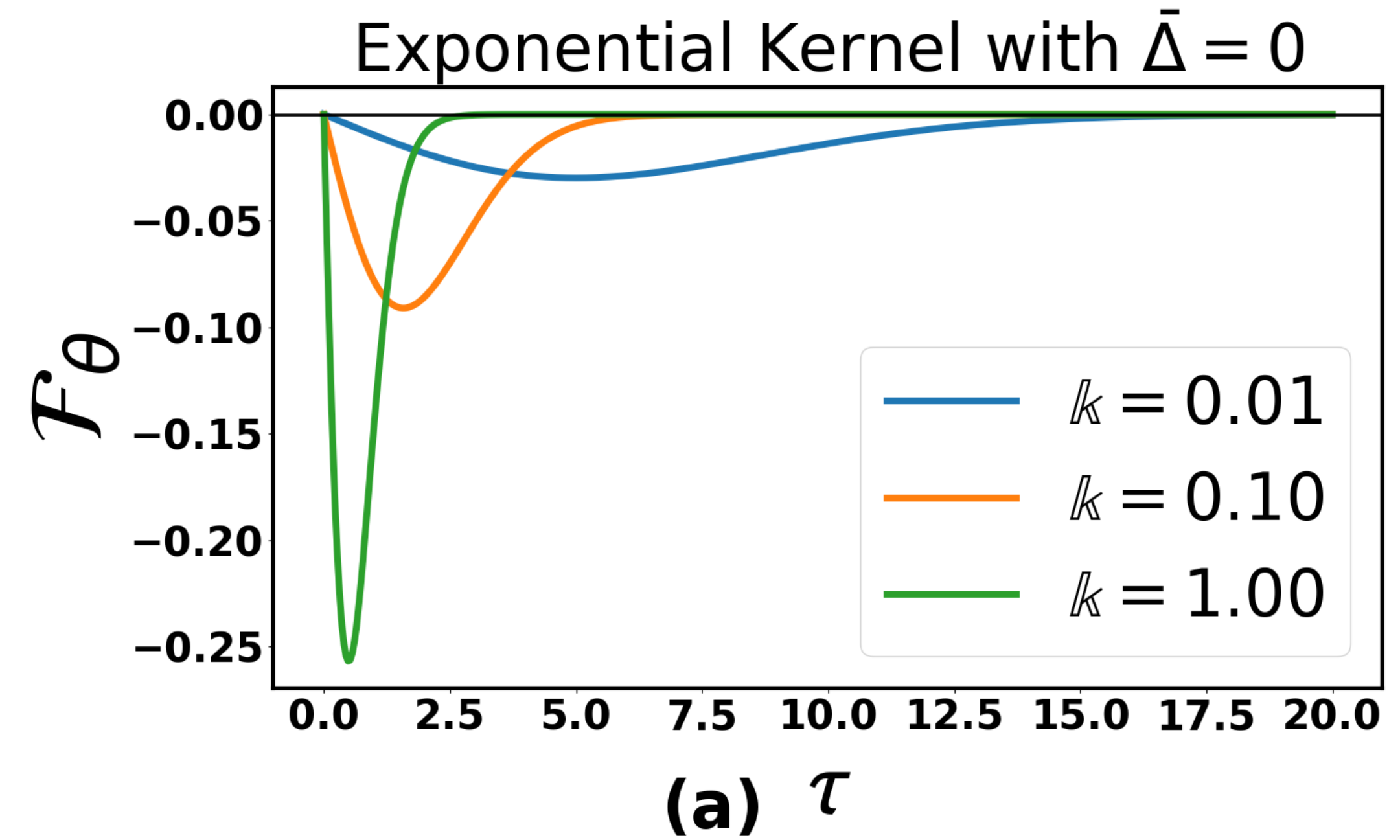}
		\includegraphics[width=4.25cm,height=4cm]{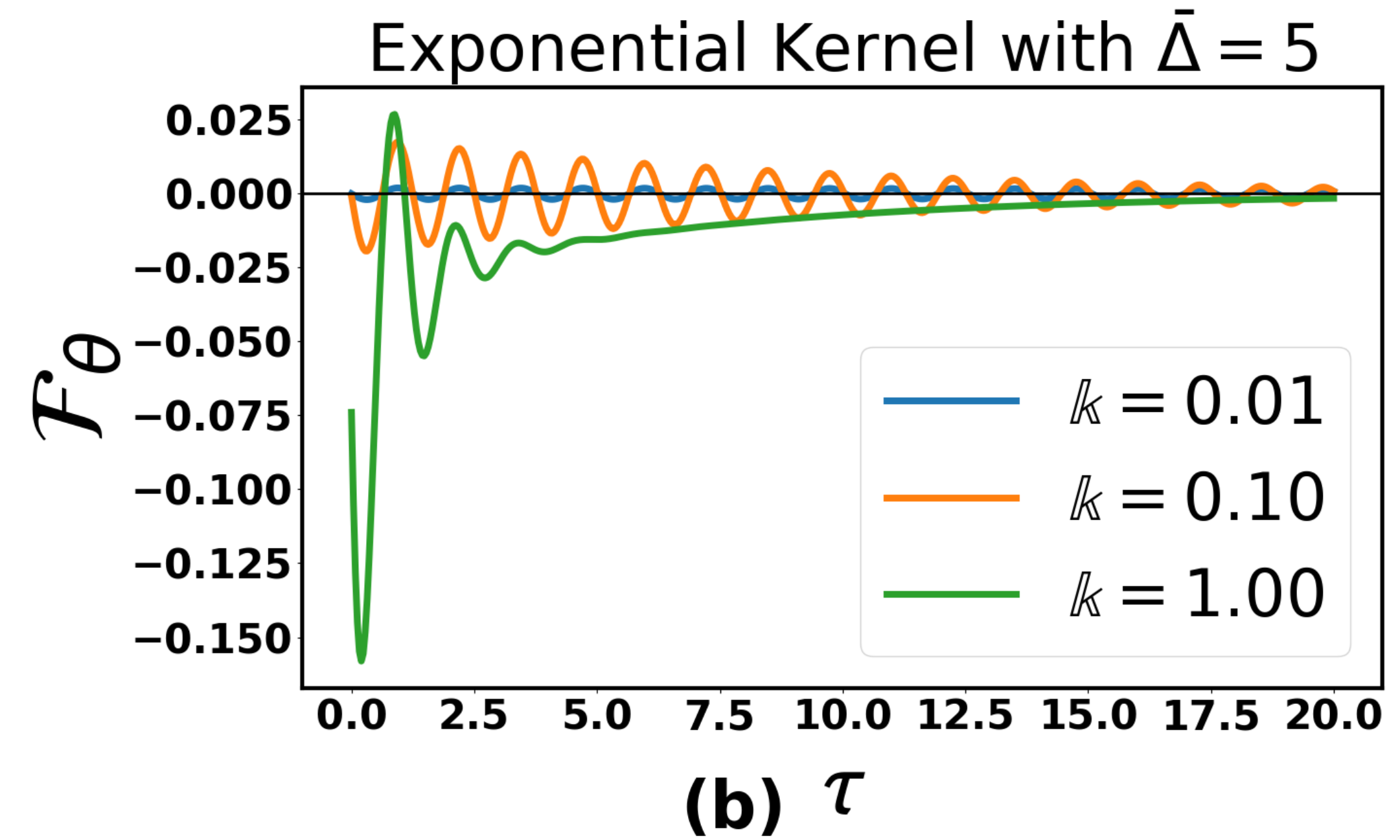}
		\includegraphics[width=4.25cm,height=4cm]{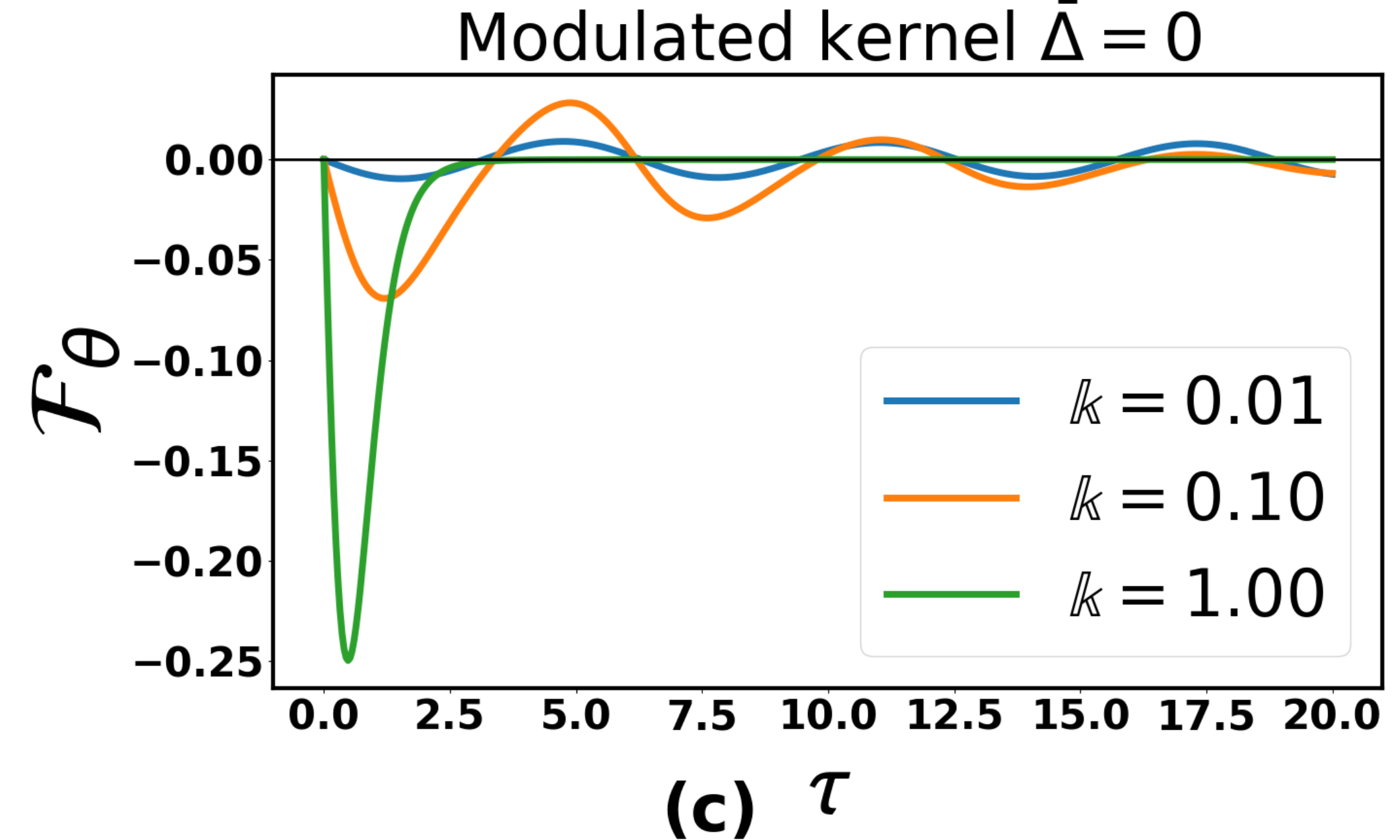}
		\includegraphics[width=4.25cm,height=4cm]{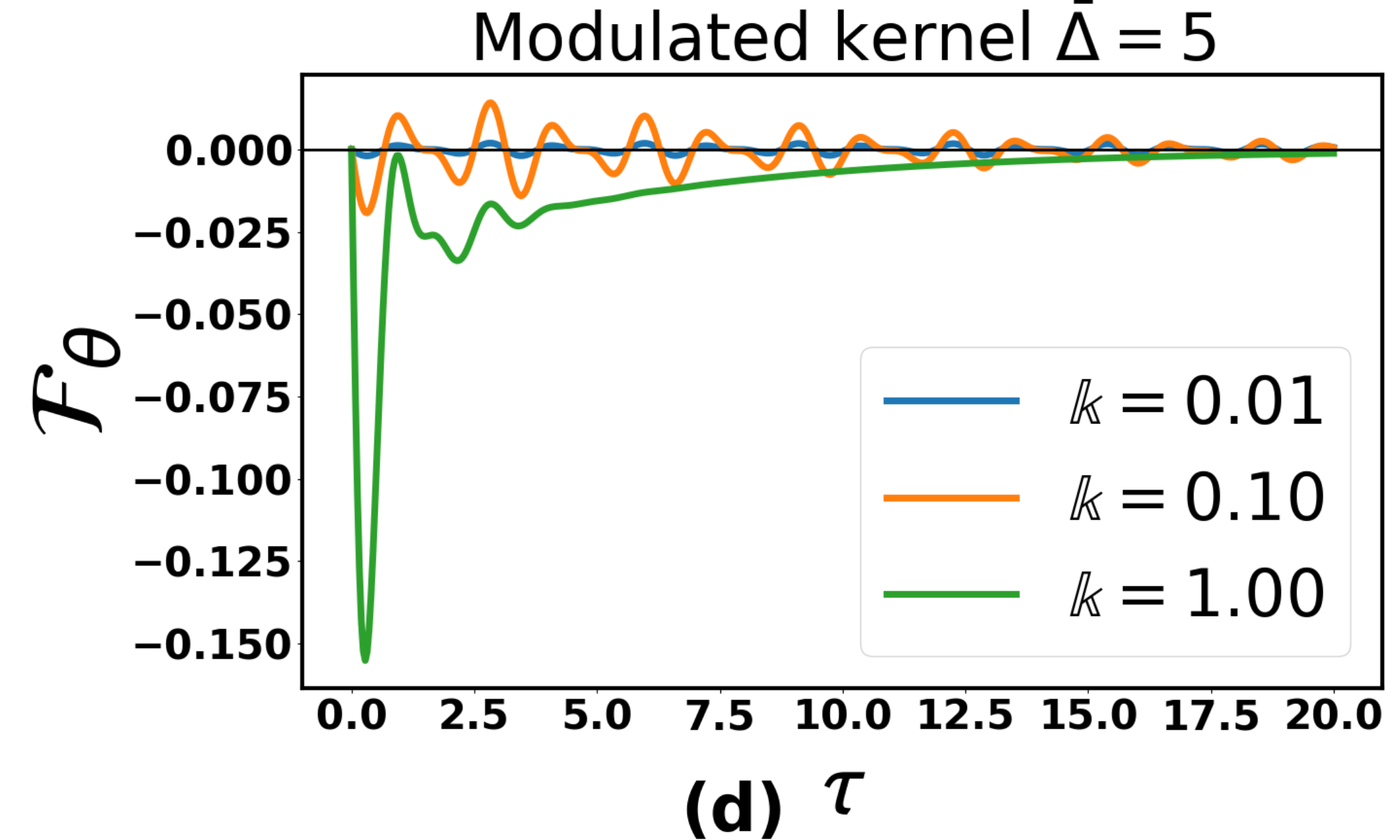}		
		\caption{Quantum Fisher information flow $\mathcal{F}_{\theta}$ for exponential and modulated memory kernels. (a) \& (b) represents the QFI flow for the exponential memory kernel while (c) \& (d) are the corresponding plots for modulated kernel. For $\bar{\Delta} =5$, we see that $\mathcal{F}_{\theta}$ captures non-Markovian behaviour for both types of memory kernels while for $\bar{\Delta}=0$, the $\mathcal{F}_{\theta}$ depicts no non-Markovian behaviour in case of exponential kernel. 
				}
		\label{QFI}
	\end{figure}

	Figure \ref{QFI}(a)-(d) illustrates the behavior of $\mathcal{F}_{\theta}$  for various values of $\bar{\Delta}$, ~$\Bbbk$ for the exponential and modulated memory kernels. In figure \ref{QFI}(a), we see that QFI flow $\mathcal{F}_{\theta}$ for $\bar{\Delta}=0$ (exponential kernel) does not have any positive values and decays to zero, thus showing no non-Markovian signature. This is also reflected in the decay of coherence and localization in figures \ref{figurea}(a) and \ref{figureb}(a). In fig. \ref{QFI}(b) for the exponential kernel with $\bar{\Delta}=5$, there are certain time intervals where $\mathcal{F}_{\theta}>0$ showing non-Markovian signature. This is also reflected in \ref{figurea}(b) and \ref{figureb}(b). The non-Markovian behaviour is also prominent in $\bar{\Delta}=0,5$ for modulated kernels as depicted in fig. \ref{QFI}(c) \& \ref{figureb}(d).
	From these plots, we observe that  $\mathcal{F}_{\theta}$ begins with a sharp negative dip, representing initial information loss or rapid decoherence, followed by oscillations that indicate the back flow of information and partial recovery, characteristic of non-Markovian behavior. The amplitude and frequency of these oscillations vary with $ \Bbbk $, demonstrating how different system parameters influence the degree of non-Markovianity. Larger values of $ \Bbbk $ lead to reduced oscillation amplitudes, suggesting weaker memory effects and faster decay of coherence. While the exponential kernel primarily controls the decay rate, the modulated kernel affects the rate and extent of these oscillations without fully suppressing the non-Markovian signatures. The persistent oscillations of $ \mathcal{F}_\theta $ confirm that significant memory effects remain in the system, indicating robust non-Markovian dynamics across a range of $ \Bbbk $ values.

	\subsection{Trace Distance as a Measure of Non-Markovianity}  
	
	Trace distance is a fundamental tool in quantum information theory, used to quantify the distinguishability between two quantum states. It plays a key role in identifying and characterizing non-Markovian effects in open quantum systems. For two quantum states \(\rho_a\) and \(\rho_b\), the trace distance is defined as \cite{3,54,63}:  
	\begin{eqnarray}  
		\mathcal{D}(\rho_a, \rho_b) = \frac{1}{2} \|\rho_a - \rho_b\| = \frac{1}{2} \text{Tr}\left[\sqrt{(\rho_a - \rho_b)^\dagger (\rho_a - \rho_b)}\right],  
	\end{eqnarray}  
	where \(\|O\| = \text{Tr}[\sqrt{O^\dagger O}]\) is the trace norm of the operator \(O\). A higher value of \(\mathcal{D}(\rho_a, \rho_b)\) indicates greater distinguishability between the states, while a value of zero implies they are indistinguishable.  
	
	In the context of quantum dynamics, the behavior of trace distance over time reveals critical information about the nature of the system's interaction with its environment.
In particular,  the trace distance decays monotonically for Markovian processes, where the system has no memory and continuously, irreversibly leaks information to the environment. In contrast, for non-Markovian dynamics, the trace distance exhibits a non-monotonic behavior over time, indicating a backflow of information from the environment into the system\cite{64,65}. This backflow is a hallmark of memory effects and distinguishes non-Markovian processes from their Markovian counterparts.

	To quantify the extent of non-Markovianity, a widely used Breuer, Laine and Pillo (BLP) measure \(\mathcal{N}(\Phi)\) has been introduced. This measure captures the strength of memory effects by evaluating the time intervals during which the trace distance increases. Formally, it is defined as \cite{7,66}:  
	\begin{eqnarray}  
		\mathcal{N}(\Phi) = \underset{\rho_a(0), \rho_b(0)}{\text{max}} \int_{\sigma(t) > 0} dt ~ \sigma(t),  
	\end{eqnarray}  
	where \(\sigma(t) = \frac{d \mathcal{D}(\rho_a, \rho_b)}{dt}\) is the rate of change of the trace distance. A positive \(\sigma(t)\) corresponds to an increase in \(\mathcal{D}(\rho_a, \rho_b)\), signaling information backflow and, therefore, non-Markovian behavior. If \(\mathcal{N}(\Phi) = 0\), the dynamics are purely Markovian, with no evidence of memory effects.  
	
	This approach to detecting non-Markovianity is particularly powerful because it does not require detailed knowledge of the underlying Hamiltonian or environmental parameters, making it highly practical for both theoretical studies and experimental implementations. By focusing on the observable dynamics of the system, the trace distance framework provides a robust and accessible way to characterize memory effects in open quantum systems, deepening our understanding of their behavior \cite{67,68}.

	
	\begin{figure}[t]
		\includegraphics[width=4cm,height=4cm]{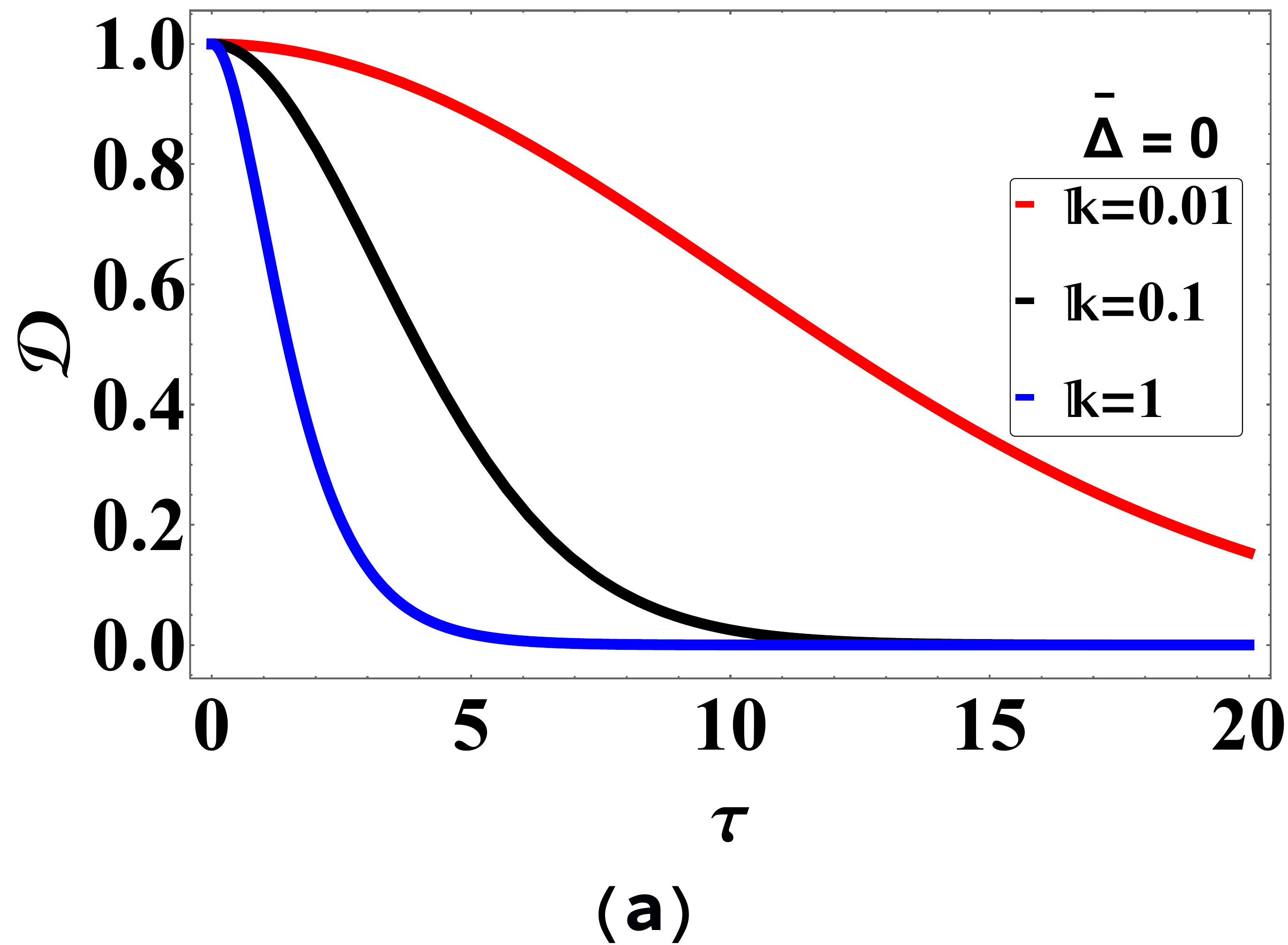}
		\includegraphics[width=4cm,height=4cm]{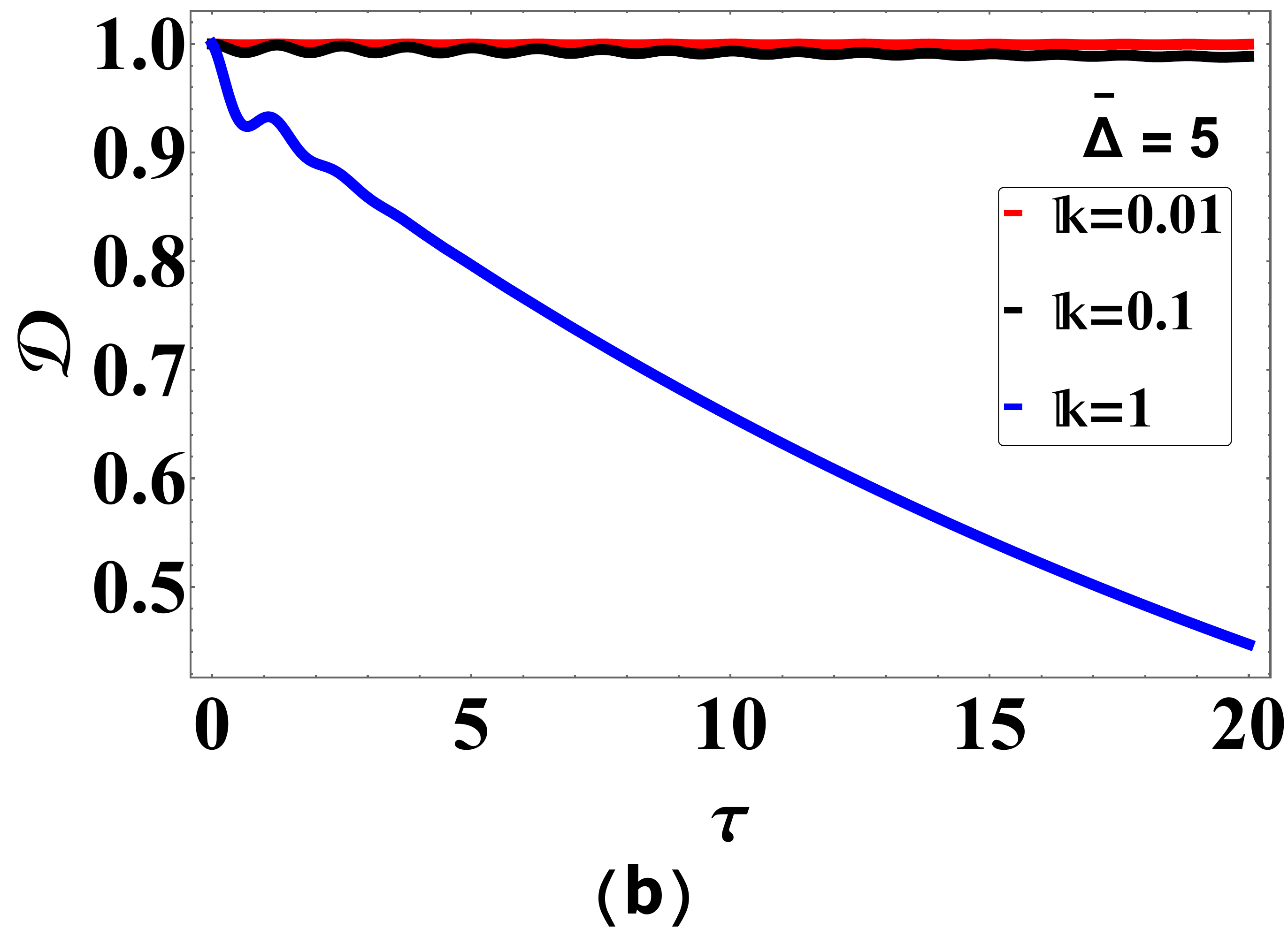}
		\includegraphics[width=4cm,height=4cm]{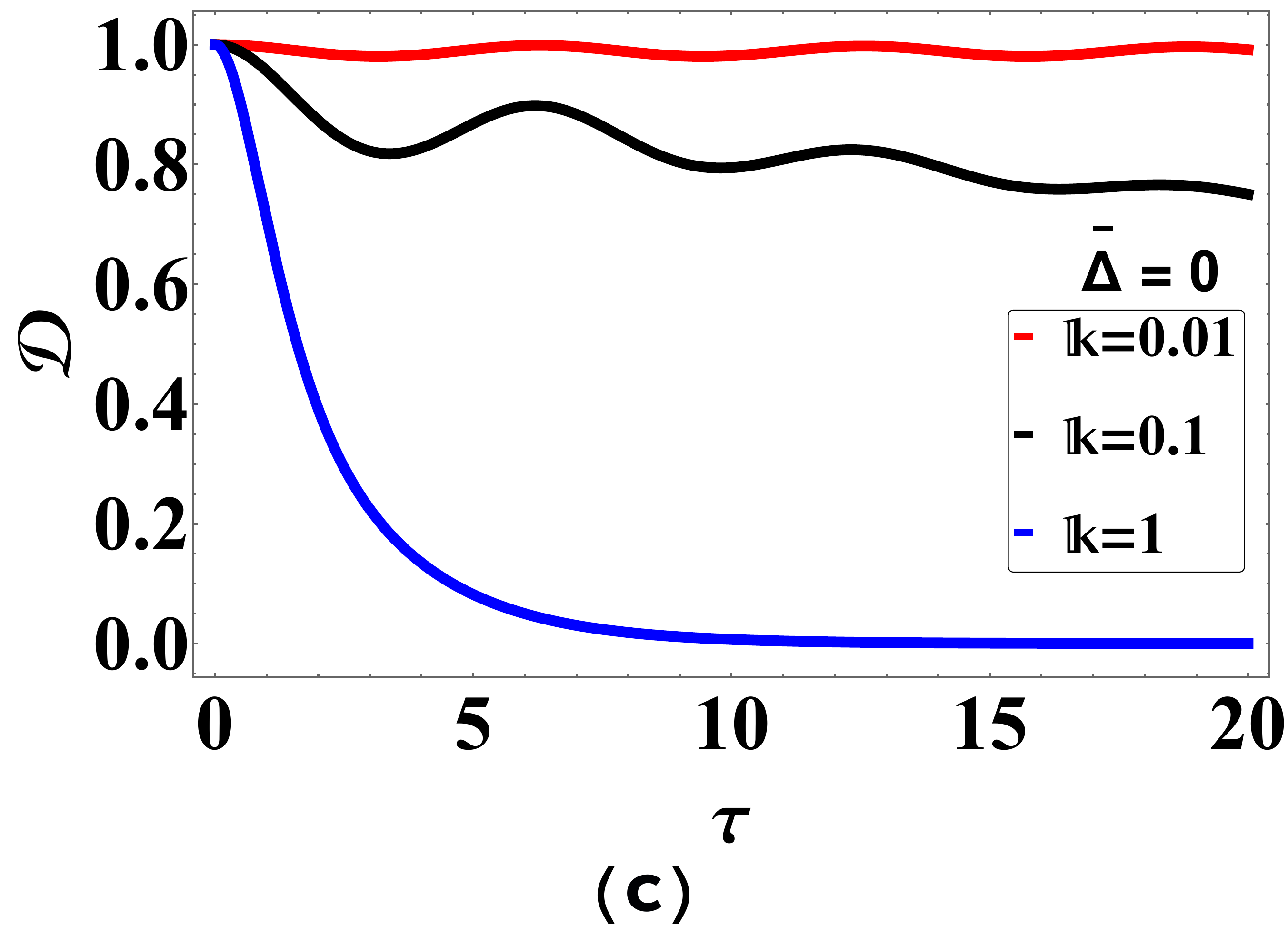}
		\includegraphics[width=4cm,height=4cm]{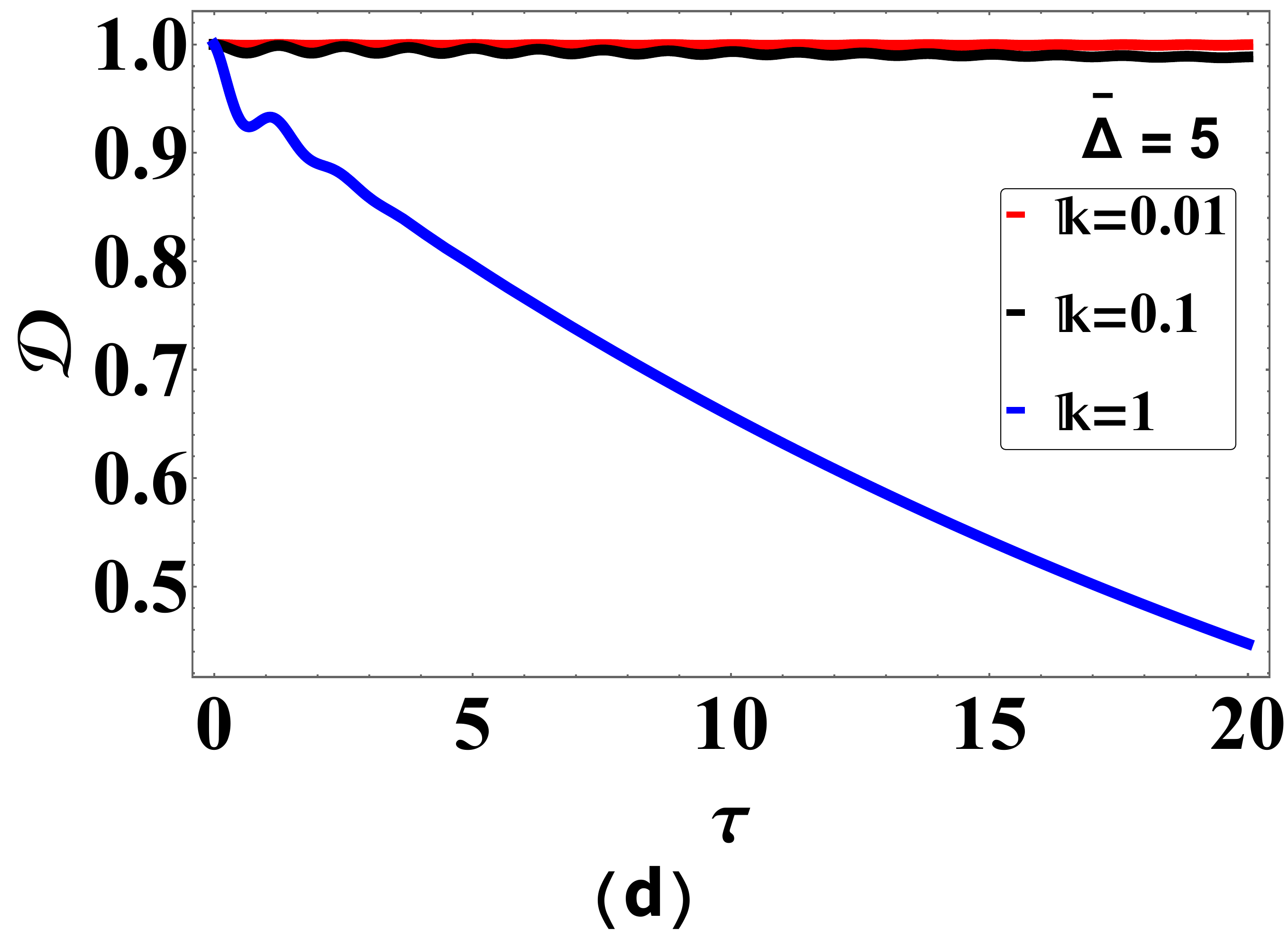}		
		\caption{Plot of trace distance $\mathcal{D}$  with respect to  $\tau$ for different parameters values. (a) and (b) represent  $\mathcal{D}$for exponential memory kernel while (c) and (d) are for modulated kernel. The parameter $\Bbbk$ is chosen in way such that the  dynamics has non-Markovian behavior. In each case, $\bar{\Delta}=0, 5 $  implying resonance and off-resonance cases respectively. }
		\label{TD}
	\end{figure}
	\begin{figure}[ht]
	\includegraphics[width=0.495\linewidth]{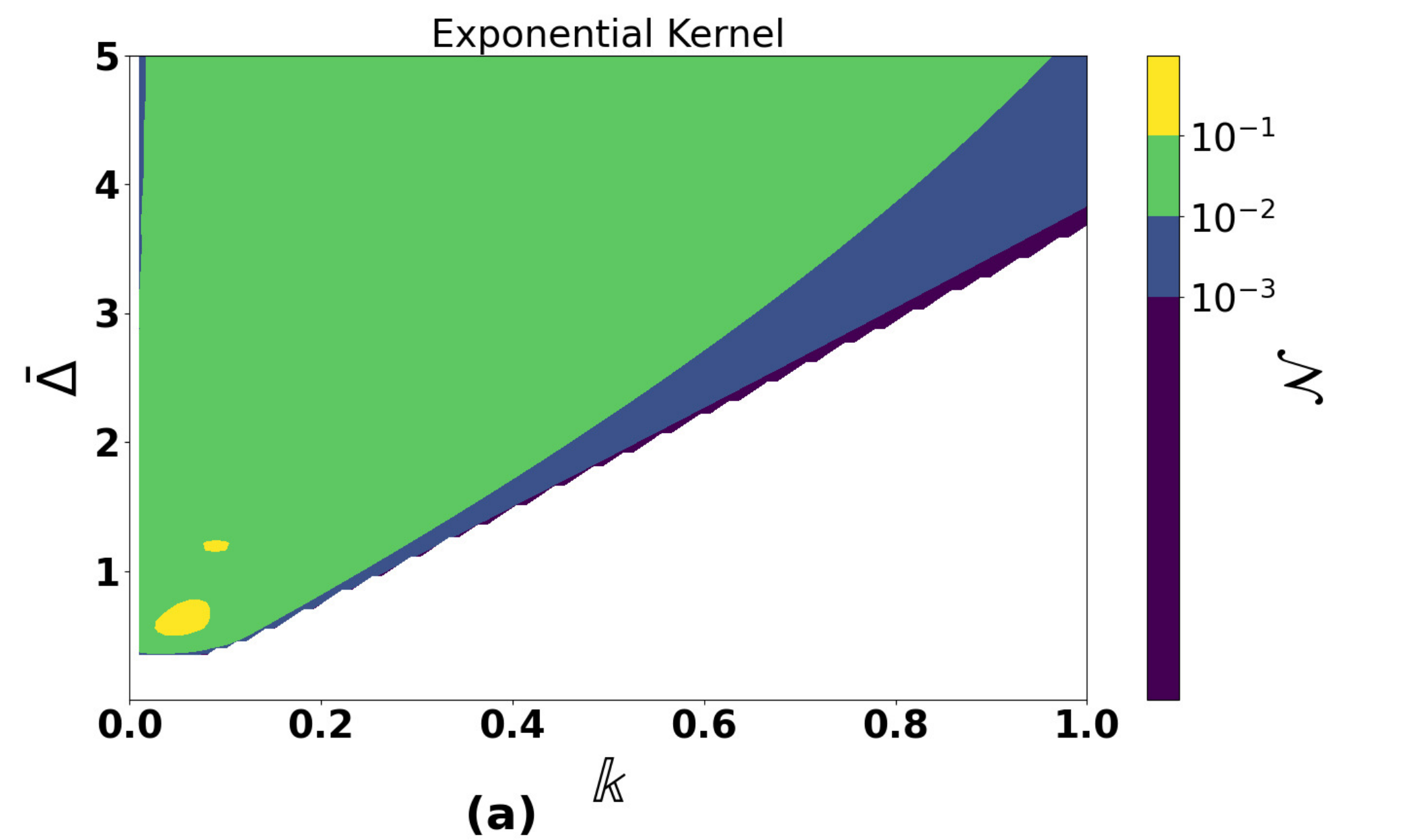} 
	\includegraphics[width=0.495\linewidth]{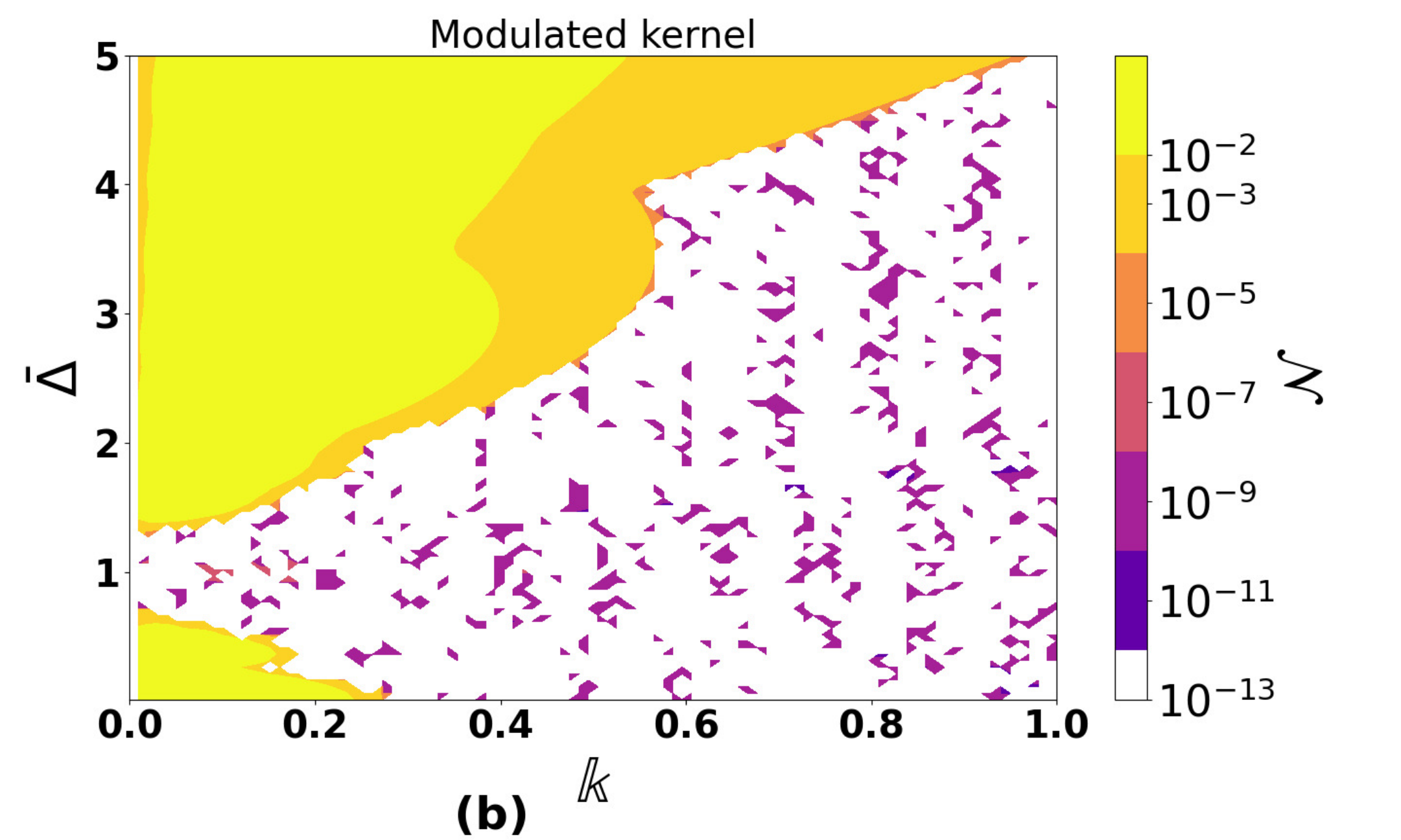}	
	\caption{{Non-Markovianity measure $\mathcal{N}(\Phi)$ based on the BLP criterion as a function of the decay rate $\Bbbk$ and detuning $\bar{\Delta}$ for (a) exponential memory kernel and (b) modulated memory kernel. The color map distinguishes three dynamical regimes: strong non-Markovianity (bright zones with pronounced revivals), weak non-Markovianity (intermediate shading with limited backflow), and partial-revival regions (sporadic or fragmented revivals). While the exponential kernel exhibits well-defined regions of memory effects, the modulated kernel produces scattered patches of non-Markovianity, reflecting irregular and frequency-dependent revivals induced by modulation.}}
		\label{ND}
	\end{figure}
	In figure \ref{TD}(a)-(d), we have plotted trace distance $\mathcal{D}$ for various values of $\bar{\Delta}$ and $\Bbbk$. In case of exponential memory kernel figure \ref{TD}(a), we observe for $\bar{\Delta}=0$, $\mathcal{D}$ goes to zero abruptly for large  $\Bbbk$ values while for $\Bbbk<<1$, it vanishes slowly. However, for $\bar{\Delta}=5$, we observe in fig \ref{TD}(b), for small  $\Bbbk$ values, $\mathcal{D}$ does not vanish abruptly but oscillates and saturates at some fixed value of coherence, thereby  maintaining coherence for longer times. The same type of behavior is observed for modulated kernel plots for trace distance $\mathcal{D}$ in figures \ref{TD}(c)-(d).  From these plots, we can infer that there exist non-Markovian features in the dynamics. This can be directly verified by the non-zero values of non-Markovian measure  $\mathcal{N}$ as shown in figures \ref{ND}(a)-(b). 
	{These plots demonstrate that non-Markovianity is highly sensitive to both the decay rate $\Bbbk$ and the detuning $\bar{\Delta}$. In the exponential kernel case [Fig.~\ref{ND}(a)], non-Markovianity decreases smoothly with increasing $\Bbbk$, while stronger memory effects appear at lower $\Bbbk$ and larger detuning values. In contrast, the modulated kernel [Fig.~\ref{ND}(b)] exhibits a richer and more irregular structure, with scattered regions of enhanced non-Markovianity. These fragmented patches signal irregular, frequency-dependent revivals of information backflow caused by the modulation component, which extends the persistence of memory effects beyond the exponential case. Together, these results reinforce our main conclusion that non-Markovianity can be selectively engineered through parameter tuning and modulation, offering practical routes for quantum control and coherence preservation in realistic quantum technologies.
		
	}

		\section{Conclusion}\label{sec:Concl}
{

		In conclusion, we investigated non-Markovian dynamics in quantum systems featuring random system–bath couplings within the versatile central spin model framework. 
		To capture complementary aspects of memory effects, we employed both the QFI flow  and the BLP measure. 
		 {The non-Markovian effects as captured by these measures for the underlying model, reflect that  QFI flow acts as a local witness, detecting instantaneous memory effects through the positivity of its time derivative, though it may miss certain non-Markovian dynamics. In contrast, BLP measure provides a global quantification of non-Markovianity by integrating optimized information backflow over the full evolution, making the two measures complementary rather than contradictory.}
		
		
	Our results demonstrate that both the degree and nature of non-Markovianity are governed by the interplay between coupling strength and the structure of the interaction kernel. Weak couplings enhance memory effects and lead to stronger revivals, whereas strong couplings suppress them, yielding nearly Markovian dynamics. Modulated interaction kernels produce qualitatively richer behavior, including frequency-sensitive enhancements of backflow.The analysis of the BLP measure under random couplings further highlights the contrast between exponential and modulated kernels. In the exponential case [Fig.~\ref{ND}(a)], non-Markovianity decreases smoothly with increasing $\Bbbk$, while stronger memory effects occur at low $\Bbbk$ and large detuning $\bar{\Delta}$. By contrast, the modulated kernel [Fig.~\ref{ND}(b)] reveals a more irregular structure, with scattered regions of enhanced non-Markovianity. These fragmented patches indicate irregular, frequency-dependent revivals of information backflow, extending the persistence of memory effects beyond what is seen in the exponential kernel. This richer dynamical structure underscores how modulation can generate non-uniform memory effects not accessible in simpler models.  
	 This complexity illustrates how kernel modulation creates non-uniform memory dynamics inaccessible to traditional exponential models.
	 
	Importantly, our findings underscore that randomness in system–bath couplings should not be viewed solely as a source of decoherence but also as a valuable resource for engineering open quantum system behavior. Disorder-induced tuning of information backflow offers practical opportunities for manipulating quantum dynamics, with potential applications in quantum metrology, communication, and computation. In particular, the ability to engineer memory effects through modulation or coupling strength optimization opens promising avenues toward noise-resilient quantum technologies.
	}

	\section*{Acknowledgments}
	The authors would like to thank Dr.~Javid A.~Naikoo (Adam Mickiewicz University, Poland) for helpful and stimulating discussions.

	

\begin{thebibliography}{}
		\bibitem{1} M. A. Nielsen and I. L. Chuang, \textit{Quantum Computation	and Quantum Information }(Cambridge University Press, 2000).
		\bibitem{2}H.-P. Breuer and F. Petruccione, \textit{The Theory of Open Quantum Systems} (Oxford 
		University Press, Oxford, 2007).
		\bibitem{3}H. P Breuer, E. M Laine, J Piilo, B Vacchini. ``Colloquium: Non-Markovian dynamics in open quantum systems." Reviews of Modern Physics {\bf 88}.2 (2016): 021002.
		\bibitem{4}Inés de Vega and D. Alonso. ``Dynamics of non-Markovian open quantum systems." Reviews of Modern Physics {\bf 89}, 015001 (2017).
		\bibitem{5}Á Rivas, SF Huelga, MB Plenio.``Quantum non-Markovianity: characterization, quantification and detection." Reports on Progress in Physics {\bf 77}.9 (2014): 094001.
		\bibitem{6} AA Budini. ``Quantum non-markovian environment-to-system backflows of information: Nonoperational vs. operational approaches." Entropy {\bf24}.5 (2022): 649.
		\bibitem{7} 	HP Breuer, EM Laine, and J  Piilo. ``Measure for the degree of non-Markovian behavior of quantum processes in open systems." Physical review letters {\bf 103}.21 (2009): 210401.
		\bibitem{8}MM Wolf, J Eisert, TS Cubitt, and JI Cirac. ``Assessing non-Markovian quantum dynamics." Physical review letters {\bf 101.}15 (2008): 150402
		\bibitem{9} S. Luo, S. Fu, and H. Song. ``Quantifying non-Markovianity via correlations." Physical Review A {\bf 86}.4 (2012): 044101.
		\bibitem{10}FF Fanchini, G Karpat, B Çakmak, LK Castelano, GH Aguilar, O Farías, SP Walborn, PH Souto Ribeiro, and MC de Oliveira. ``Non-Markovianity through accessible information." Physical Review Letters {\bf 112}.21 (2014): 210402.
		\bibitem{11}H Song, S Luo, Y Hong. ``Quantum non-Markovianity based on the Fisher-information matrix." Physical Review A {\bf 91}.4 (2015): 042110.
		\bibitem{12} P Abiuso, M Scandi, D De Santis, J Surace . ``Characterizing (non-) Markovianity through Fisher information." SciPost Physics {\bf 15}.1 (2023): 014.
		\bibitem{14}Ł Cywiński, WM Witzel, and S Das Sarma. "Pure quantum dephasing of a solid-state electron spin qubit in a large nuclear spin bath coupled by long-range hyperfine-mediated interactions." Physical Review B—Condensed Matter and Materials Physics 79.24 (2009): 245314.
		\bibitem{15}P. Cappellaro, L. Jiang, J. S. Hodges, and M. D. Lukin. "Coherence and control of quantum registers based on electronic spin in a nuclear spin bath." Physical review letters 102.21 (2009): 210502.
		\bibitem{16}JM Taylor, CM Marcus and MD Lukin. "Long-lived memory for mesoscopic quantum bits." Physical review letters 90.20 (2003): 206803.
		\bibitem{17}V. V. Dobrovitski, A. E. Feiguin, D. D. Awschalom, and R. Hansonl. "Decoherence dynamics of a single spin versus spin ensemble." Physical Review B—Condensed Matter and Materials Physics 77.24 (2008): 245212.
		\bibitem{18}J Jing, and LA Wu. "Decoherence and control of a qubit in spin baths: an exact master equation study." Scientific reports 8.1 (2018): 1471.
		\bibitem{19}	L. Cywi´nski, Dephasing of electron spin qubits due to
		their interaction with nuclei in quantum dots, Acta
		Phys. Pol. A 119, 576 (2011).
		\bibitem{20} B. Urbaszek, X. Marie, T. Amand, O. Krebs, P. Voisin,
		P. Maletinsky, A. H¨ogele, and A. Imamoglu, Nuclear
		spin physics in quantum dots: An optical investigation,
		Rev. Mod. Phys. 85, 79 (2013).
		\bibitem{21} R. van den Berg, G. P. Brandino, O. El Araby, R. M.
		Konik, V. Gritsev, and J.-S. Caux, Competing interactions in semiconductor quantum dots, Phys. Rev. B 90,
		155117 (2014).
		\bibitem{22} W. Yang, W.-L. Ma, and R.-B. Liu, Quantum manybody theory for electron spin decoherence in nanoscale
		nuclear spin baths, Rep. Prog. Phys. 80, 016001 (2016).
		
		\bibitem{23}  A. Khaetskii, D. Loss, and L. Glazman, Electron spin
		evolution induced by interaction with nuclei in a quantum dot, Phys. Rev. B 67, 195329 (2003).
		\bibitem{24}  J. Schliemann, A. Khaetskii, and D. Loss, Electron spin
		dynamics in quantum dots and related nanostructures
		due to hyperfine interaction with nuclei, J. Phys.: Condens. Matter 15, R1809 (2003).
		\bibitem{25} W. A. Coish and D. Loss, Hyperfine interaction in a
		quantum dot: Non-markovian electron spin dynamics,
		Phys. Rev. B 70, 195340 (2004).
		\bibitem{26}  C. Deng and X. Hu, Analytical solution of electron spin
		decoherence through hyperfine interaction in a quantum
		dot, Phys. Rev. B 73, 241303 (2006).
		\bibitem{27}  R. Hanson, L. P. Kouwenhoven, J. R. Petta, S. Tarucha,
		and L. M. K. Vandersypen, Spins in few-electron quantum dots, Rev. Mod. Phys. 79, 1217 (2007).
		
		
		\bibitem{nv1}Erik Bauch, Swati Singh, Junghyun Lee, Connor A. Hart, Jennifer M. Schloss, Matthew J. Turner, John F. Barry, Linh M. Pham, Nir Bar-Gill, Susanne F. Yelin, and Ronald L. Walsworth. "Decoherence of ensembles of nitrogen-vacancy centers in diamond." Physical Review B 102.13 (2020): 134210.
		\bibitem{nv2} A. Laraoui, J. S. Hodges, C. A. Ryan, and C. A. Meriles,
		Diamond nitrogen-vacancy center as a probe of random
		fluctuations in a nuclear spin ensemble, Phys. Rev. B
		84, 104301 (2011).
		\bibitem{nv3} N. Zhao, S.-W. Ho, and R.-B. Liu, Decoherence and
		dynamical decoupling control of nitrogen vacancy center
		electron spins in nuclear spin baths, Phys. Rev. B 85,
		115303 (2012).
		\bibitem{nv4} L. T. Hall, J. H. Cole, and L. C. L. Hollenberg, Analytic solutions to the central-spin problem for nitrogenvacancy centers in diamond, Phys. Rev. B 90, 075201
		(2014).
		\bibitem{nv5} I. Schwartz, J. Scheuer, B. Tratzmiller, S. M¨uller,
		Q. Chen, I. Dhand, Z.-Y. Wang, C. M¨uller, B. Naydenov, F. Jelezko, and M. B. Plenio, Robust optical polarization of nuclear spin baths using hamiltonian engineering of nitrogen-vacancy center quantum dynamics,
		Sci. Adv. 4, eaat8978 (2018).
		\bibitem{36}M Onizhuk, YX Wang, J Nagura, AA Clerk, G Galli . "Understanding central spin decoherence due to interacting dissipative spin baths." Physical Review Letters 132.25 (2024): 250401.
		\bibitem{37} M.-H. Yung, Spin star as a switch for quantum networks, J. Phys. B 44, 135504 (2011).
		\bibitem{38} H.-L. Deng and X.-M. Fang, Quantum phase transitions
		and quantum communication in a spin star system, J.
		Phys. B 41, 025503 (2008).
		\bibitem{39} M. C. Tran and J. M. Taylor, Blind quantum
		computation using the central spin hamiltonian,
		arXiv:1801.04006 (2018).
		
		\bibitem{40} G. Anikeeva, O. Markovi´c, V. Borish, J. A. Hines, S. V.
		Rajagopal, E. S. Cooper, A. Periwal, A. Safavi-Naeini,
		E. J. Davis, and M. Schleier-Smith, Number partitioning with grover’s algorithm in central spin systems, PRX
		Quantum 2, 020319 (2021).
		
		
		\bibitem{41}JH Wilson, BM. Fregoso, and VM. Galitski. Entanglement dynamics in a non-Markovian environment: An exactly solvable model. Physical Review B—Condensed Matter and Materials Physics 85.17 (2012): 174304.
		\bibitem{42}M Bortz, and J Stolze. "Spin and entanglement dynamics in the central-spin model with homogeneous couplings." Journal of Statistical Mechanics: Theory and Experiment 2007.06 (2007): P06018.
		\bibitem{43}R Nepomechie and XW Guan. ``The spin-s homogeneous central spin model: exact spectrum and dynamics." Journal of Statistical Mechanics: Theory and Experiment 2018.10 (2018): 103104.
		
		\bibitem{45}Villazon, Tamiro, Anushya Chandran, and Pieter W. Claeys. "Integrability and dark states in an anisotropic central spin model." Physical Review Research 2.3 (2020): 032052.
		\bibitem{46}Villazon, T., Claeys, P.W., Pandey, M. et al. Persistent dark states in anisotropic central spin models. Sci Rep 10, 16080 (2020).
		\bibitem{47}Khurshudyan, Martiros. "Controlled State Transfer in Central Spin Models." Symmetry 16.4 (2024): 489.
		\bibitem{44}Dobrzyniecki, Jacek, and Michał Tomza. "Quantum simulation of the central spin model with a Rydberg atom and polar molecules in optical tweezers." Physical Review A 108.5 (2023): 052618
		\bibitem{Gross}C Gross, I Bloch. ``Quantum simulations with ultracold atoms in optical lattices." Science 357.6355 (2017): 995-1001.
		\bibitem{AP}A Prakash, B Hebbe Madhusudhana. ``Characterizing non-Markovian and coherent errors in quantum simulation." Physical Review Research 6.4 (2024): 043127.
		\bibitem{48}Dhara, Prajit, and Saikat Guha. "Phonon-induced decoherence in color-center qubits." Physical Review Research 6.1 (2024): 013055.
		\bibitem{49}Gerasimov, L. V., et al. "Coupled dynamics of spin qubits in optical dipole microtraps: Application to the error analysis of a Rydberg-blockade gate." Physical Review A 106.4 (2022): 042410.
		\bibitem{jw}E. Fradkin. ``Jordan-Wigner transformation for quantum-spin systems in two dimensions and fractional statistics." Physical review letters 63.3 (1989): 322.
		\bibitem{50}S Alipour, M. Mehboudi, and AT. Rezakhani. ``Quantum metrology in open systems: dissipative Cramér-Rao bound." Physical review letters {\bf 112}.12 (2014): 120405.
		\bibitem{51}J Naikoo, WC Ravindra, and J Kołodyński. ``Multiparameter estimation perspective on non-Hermitian singularity-enhanced sensing." Physical Review Letters {\bf 131}.22 (2023): 220801.
		\bibitem{52} F. Chapeau-Blondeau. ``Optimizing qubit phase estimation." Physical Review A {\bf 94}.2 (2016): 022334.
		\bibitem{53} F. Chapeau-Blondeau. ``Optimized probing states for qubit phase estimation with general quantum noise." Physical Review A {\bf 91}.5 (2015): 052310.	
		\bibitem{54}S. Haseli, G. Karpat, S. Salimi, AS. Khorashad, F F. Fanchini, BC¸ akmak, GH. Aguilar,
		SP. Walborn and P. H. Souto Ribeiro.``Non-Markovianity through flow of information between a system and an environment." Physical Review A {\bf 90}.5 (2014): 052118.
		\bibitem{55}XM Lu, X Wang, and CP Sun. ``Quantum Fisher information flow and non-Markovian processes of open systems." Physical Review A {\bf 82}.4 (2010): 042103.
		
		\bibitem{56} Y. M. Zhang, X. W. Li, W. Yang, and G. R. Jin. ``Quantum Fisher information of entangled coherent states in the presence of photon loss." Physical Review A {\bf 88}.4 (2013): 043832.
		\bibitem{57}J Joo, WJ Munro, and TP Spiller. ``Quantum metrology with entangled coherent states." Physical review letters {\bf 107}.8 (2011): 083601.
		
		\bibitem{58}J. Naikoo, S. Dutta, and S. Banerjee. ``Facets of quantum information under non-Markovian evolution." Physical Review A {\bf 99}.4 (2019): 042128.
		\bibitem{59} XY Chen, NN Zhang , WT He, XY Kong, MJ Tao, FG Deng, Q Ai  and GL Long.``Global correlation and local information flows in controllable non-Markovian open quantum dynamics." npj Quantum Information {\bf 8}.1 (2022): 22.
		\bibitem{60}A. Rivas, S. F. Huelga, and M. B. Plenio.  ``Entanglement and non-Markovianity of quantum evolutions." Physical review letters {\bf 105}.5 (2010): 050403.
		\bibitem{61}B Vacchini, A Smirne, EM Laine, J Piilo``Markovianity and non-Markovianity in quantum and classical systems." New Journal of Physics {\bf 13}.9 (2011): 093004.
		
		\bibitem{63}B Bylicka, M Johansson, A Acin. ``Constructive method for detecting the information backflow of non-Markovian dynamics." Physical review letters {\bf 118}.12 (2017): 120501.
		\bibitem{64} E.-M. Laine, J. Piilo, and H.P. Breuer. ``Measure for the non-Markovianity of quantum processes." Physical Review A {\bf 81}.6 (2010): 062115.
		\bibitem{65} S Wißmann, A Karlsson, EM Laine, J Piilo, HP Breuer. ``Optimal state pairs for non-Markovian quantum dynamics." Physical Review A {\bf 86}.6 (2012): 062108.
		\bibitem{66}G Amato, HP Breuer, B Vacchini. ``Generalized trace distance approach to quantum non-Markovianity and detection of initial correlations." Physical Review A {\bf 98}.1 (2018): 012120.
		
		\bibitem{67}J Łuczka. ``Spin in contact with thermostat: Exact reduced dynamics." Physica A: Statistical Mechanics and its Applications {\bf 167}.3 (1990): 919-934..
		\bibitem{68}AJ Leggett, S Chakravarty, AT Dorsey, MPA  Fisher, A Garg, and W Zwerger ``Dynamics of the dissipative two-state system." Reviews of Modern Physics {\bf 59}.1 (1987): 1.
		\bibitem{69} Meng X, Sun Y, Wang Q, Ren J, Cai X, Czerwinski A. ``Dephasing dynamics in a non-equilibrium fluctuating environment." Entropy {\bf 25}.4 (2023): 634.
		\bibitem{70}J. W Negele, \textit{Quantum Many-particle Systems} (1st ed. 1998). CRC Press.
		\bibitem{Assa}A Auerbach,\textit{ Interacting electrons and quantum magnetism.} Springer Science \& Business Media, 2012.
		\bibitem{Col}P Coleman,\textit{Introduction to many-body physics}. Cambridge University Press, 2015.
	\end{thebibliography}

	\appendix
	\setcounter{secnumdepth}{3} 
	\section{Jordan Wigner transformation}\label{apn:JWT}

{ Corresponding to the spin degree of freedom, for each lattice site  $k$, we can define the corresponding observables $S_{k}^{x}, S_{k}^{x}, S_{k}^{x}$\cite{Assa,Col}.  In order to relate such spin system to a fermion system, it is convinient to introduce the raising a lowering operators $	S_{k}^{\pm} =  S_{k}^{x} \pm i S_{k}^{y}$. These operators raise and lower the spins as $S_{k}^{+} |\downarrow \rangle = |\uparrow \rangle \quad S_{k}^{-} |\downarrow \rangle = 0 \quad S_{k}^{+} |\uparrow \rangle =0  \quad S_{k}^{-} |\uparrow \rangle = |\downarrow \rangle $ . These following properties are important to note in order to move from spin operators to fermions 
	\begin{align}
		\{ S_{k}^{+}, S_{k}^{-} \} &= S_{k}^{+} S_{k}^{-} + S_{k}^{-} S_{k}^{+} = 1  \\&{\rm same~site~anticommutation~relation} \nonumber \\
		[S_{j}^{\pm}, S_{k}^{\pm} ] &= 0 \quad j\ne k \\&{\rm different~site~commutation~relation} \nonumber \\
		(S_{k}^{\pm})^2 &= 0 \\ 
		S_{k}^{z} &= S_{k}^{+} S_{k}^{-} - \frac{1}{2}.
	\end{align}
	Now, fermions have creation and annihilation operators (\(f_k^\dagger, f_k\)) that function similarly to raising and lowering operators. {Specifically:~~$
	f_{k}^\dagger |\downarrow \rangle = |\uparrow \rangle, \quad f_{k} |\downarrow \rangle = 0, \quad f_{k}^\dagger |\uparrow \rangle = 0, \quad f_{k} |\uparrow\rangle = |\downarrow \rangle$,
	and they satisfy the same-site anticommutation relation:
	\begin{align}
		\{f_k, f_k^\dagger\} &= 1, \label{eq:anticomm1}
	\end{align}}
	This behavior is identical to the raising and lowering operators in spin systems. The number operator is defined as \(n_k = f_k^\dagger f_k\), which gives the number of fermions in a particular state associated with lattice site \(k\).
	
	Fermions obey anticommutation relations not only with operators at the same site but also across different lattice sites. This means they satisfy the following anticommutation relations for any two sites \(j\) and \(k\):
{	
	\begin{align}
		\{f_j, f_k\} &= \{f_j^\dagger, f_k^\dagger\} = \{f_j, f_k^\dagger\} = 0 \quad j\ne k  \label{eq:anticomm2}
	\end{align}}
	Despite this difference, Jordan and Wigner in 1928 \cite{jw}, showed that the spin operators for an entire chain can be represented exactly in terms of fermion operators through the following mapping:
	\begin{eqnarray}
		S_k^{-} &= \exp(+ i \pi \sum_{l=1}^{k-1} f_l^\dagger f_l )f_k  \label{eq:jw1} \\
		S_k^{+} &= f_k^\dagger \exp(-i \pi \sum_{l=1}^{k-1} f_l^\dagger f_l ) \label{eq:jw2}
	\end{eqnarray} where the sum runs over all fermion sites from  $l = 1 $ to $k-1$, and k is the index of the spin operator being mapped.
	From these definitions, we can write : 
	\begin{eqnarray}
		S_k^z = f_k^\dagger f_k - 1/2 \label{eq:jw3}
	\end{eqnarray}
	These three relations (\ref{eq:jw1},\ref{eq:jw2} and \ref{eq:jw3}) together define the Jordan-Wigner transformation.}
	
	\section{Interaction Hamiltonian in Eq.  (\ref{eq:Hint}) of the main text}\label{apn:Hint}
	{ The interaction picture of an operator is defined with respect to the free part of the total Hamiltonian, which in our case is  $\mathbf{H}_{S} + \textbf{H}_{B}$ as given in Eq.~(\ref{eq:HSHB}). For some operator $\mathcal{O}$, we have  $\mathcal{O}_{int}(t)= e^{i (\mathbf{H}_{S} + \textbf{H}_{B}) t} \mathcal{O} e^{- i (\mathbf{H}_{S} + \textbf{H}_{B}) t}$. In particular, if $\mathcal{O} = \sum_{j} S_{j} \otimes B_{j}$ where  operators $S_{j}$ and $B_{j}$ are associated with  system and bath, respectively, then $\mathcal{O}_{int}(t) = \sum_{j} \left(e^{i \mathbf{H}_{S} t} S_{j} e^{- i \mathbf{H}_{S} t}  \otimes  e^{i \mathbf{H}_{B} t} B_{j} e^{- i \mathbf{H}_{B} t}\right)$.  
		{
\begin{align}
	\mathbf{H}_{int}(t) &= e^{i (\mathbf{H}_{S} + \mathbf{H}_{B}) t} \mathbf{H}_I e^{-i (\mathbf{H}_{S} + \mathbf{H}_{B}) t}, \nonumber \\
	&= \frac{1}{\sqrt{N}} \sum_k J_k(t) \Big[ 
	\sigma_0^+ \cdot \hat{f}_k 
	+ \sigma_0^- \cdot \hat{f}^{\dagger}_k 
	+ 2 \sigma_0^z \cdot \hat{f}^{\dagger}_k \hat{f}_k 
	\Big], \nonumber \\
	&= \frac{1}{\sqrt{N}} \sum_k J_k(t) \Big[ 
	e^{i \frac{\omega_r t}{2} \sigma_0^z} \sigma_0^+ e^{-i \frac{\omega_r t}{2} \sigma_0^z} 
	\cdot e^{i \sum_{j=1}^N \omega_j \hat{f}^{\dagger}_j \hat{f}_j t} \hat{f}_k  \times e^{-i \sum_{j=1}^N \omega_j \hat{f}^{\dagger}_j \hat{f}_j t} \nonumber \\
	&\quad + e^{i \frac{\omega_r t}{2} \sigma_0^z} \sigma_0^- e^{-i \frac{\omega_r t}{2} \sigma_0^z} 
	\cdot e^{i \sum_{j=1}^N \omega_j \hat{f}^{\dagger}_j \hat{f}_j t} \hat{f}_k^{\dagger}\times e^{-i \sum_{j=1}^N \omega_j \hat{f}^{\dagger}_j \hat{f}_j t}+ 2 e^{i \frac{\omega_r t}{2} \sigma_0^z} \sigma_0^z e^{-i \frac{\omega_r t}{2} \sigma_0^z} 
	\cdot e^{i \sum_{j=1}^N \omega_j \hat{f}^{\dagger}_j \hat{f}_j t} \nonumber \\
	&\quad \times \hat{f}_k^{\dagger} \hat{f}_k 
	e^{-i \sum_{j=1}^N \omega_j \hat{f}^{\dagger}_j \hat{f}_j t} 
	\Big].
\end{align}
Using standard operator rotation identities under unitary evolution, we have the following transformations:
	
	\begin{align*}
		e^{i \frac{\omega_r t}{2} \sigma_0^z} \, \sigma_0^+ \, e^{-i \frac{\omega_r t}{2} \sigma_0^z} &= \sigma_0^+ \, e^{i \omega_r t} ,~~
		e^{i \frac{\omega_r t}{2} \sigma_0^z} \, \sigma_0^- \, e^{-i \frac{\omega_r t}{2} \sigma_0^z} = \sigma_0^- \, e^{-i \omega_r t} 
	\end{align*}
	Similarly, for the fermion bath operators:
	\begin{align*}
		e^{i \sum_k \omega_k f_k^\dagger f_k t} \, f_k \, e^{-i \sum_k \omega_k f_k^\dagger f_k t} &= f_k \, e^{-i \omega_k t}  \\
		e^{i \sum_k \omega_k f_k^\dagger f_k t} \, f_k^\dagger \, e^{-i \sum_k \omega_k f_k^\dagger f_k t} &= f_k^\dagger \, e^{i \omega_k t}  \\
		e^{i \sum_k \omega_k f_k^\dagger f_k t} \, f_k^\dagger f_k \, e^{-i \sum_k \omega_k f_k^\dagger f_k t} &= f_k^\dagger f_k 
\end{align*}}

	\begin{align}
		\mathbf{H}_{int}(t) &= \frac{1}{\sqrt{N}} \sum_k J_k(t) \bigg[ 
		\sigma^+ f_k e^{i (\omega_r - \omega_k)t}
		+ \sigma^- f^{\dagger}_k e^{-i (\omega_r - \omega_k)t} + 2\sigma^z f^{\dagger}_k f_k 
		\bigg].
		\label{eq:Hint_split}
	\end{align}	
	\section{Derivation of Lindblad form in Eq. (\ref{MAS1}) of the main text }\label{apn:LindbladEqn}
	From equation(\ref{eq:ME}), we  have 
\begin{eqnarray}
	\dot{\rho}^S_{int}(t) &=& - \int_0^t dt' \Big[ 
	{\rm Tr_B} \big\{ \mathbf{H}_{int}(t) \mathbf{H}_{int}(t') \rho^S_{int}(t) \otimes |0_B\rangle\langle 0_B| \big\} - {\rm Tr_B} \big\{ \mathbf{H}_{int}(t) \rho^S_{int}(t) \otimes |0_B\rangle\langle 0_B| \mathbf{H}_{int}(t') \big\} \nonumber \\
	&& \quad - {\rm Tr_B} \big\{ \mathbf{H}_{int}(t') \rho^S_{int}(t) \otimes |0_B\rangle\langle 0_B| \mathbf{H}_{int}(t) \big\}  + {\rm Tr_B} \big\{ \rho^S_{int}(t) \otimes |0_B\rangle\langle 0_B| \mathbf{H}_{int}(t') \mathbf{H}_{int}(t) \big\} \Big].
	\label{mas1}
\end{eqnarray}

	In the interaction picture, we have  $\mathbf{H}_{int}(t) = \frac{1}{\sqrt{N}}\sum_k J_k(t)[ \sigma^+ f_k e^{i (\omega_0 - \omega_k)t}+ \sigma^- f^{\dagger}_k e^{-i (\omega_0 - \omega_k)t} + 2\sigma^z f^{\dagger}_k f_k ]$. Now from equation (\ref{mas1}), the first term can be written as :

\begin{align}
	& \mathrm{Tr}_B \big\{ \mathbf{H}_{\mathrm{int}}(t) \, \mathbf{H}_{\mathrm{int}}(t') \, \rho^S_{\mathrm{int}}(t) \otimes |0_B\rangle\langle 0_B| \big\} \nonumber \\
	&= \frac{1}{N} \sum_{k,l} J_k(t) J_l(t') \, \mathrm{Tr}_B \Big[ 
	\big( \sigma^+ f_k \, e^{i (\omega_0 - \omega_k) t} 
	+ \sigma^- f_k^\dagger \, e^{-i (\omega_0 - \omega_k) t} 
	+ 2 \sigma^z f_k^\dagger f_k \big) \nonumber \\
	&\qquad \times 
	\big( \sigma^+ f_l \, e^{i (\omega_0 - \omega_l) t'} 
	+ \sigma^- f_l^\dagger \, e^{-i (\omega_0 - \omega_l) t'} 
	+ 2 \sigma^z f_l^\dagger f_l \big) \,
	\rho^S_{\mathrm{int}}(t) \otimes |0_B\rangle\langle 0_B| \Big] \nonumber \\
	&= \frac{1}{N} \sum_{k,l} J_k(t) J_l(t') \, \Big[ 
	\; \sigma^+ \sigma^+ \langle f_k f_l \rangle 
	e^{i (\omega_0 - \omega_k)t} e^{i (\omega_0 - \omega_l)t'} \quad + \sigma^+ \sigma^- \langle f_k f_l^\dagger \rangle 
	e^{i (\omega_0 - \omega_k)t} e^{-i (\omega_0 - \omega_l)t'} \nonumber \\
	&\quad + 2 \sigma^+ \sigma^z \langle f_k f_l^\dagger f_l \rangle 
	e^{i (\omega_0 - \omega_k)t} \quad + \sigma^- \sigma^+ \langle f_k^\dagger f_l \rangle 
	e^{-i (\omega_0 - \omega_k)t} e^{i (\omega_0 - \omega_l)t'} \quad + \sigma^- \sigma^- \langle f_k^\dagger f_l^\dagger \rangle 
	e^{-i (\omega_0 - \omega_k)t} e^{-i (\omega_0 - \omega_l)t'} \nonumber \\
	&\quad + 2 \sigma^- \sigma^z \langle f_k^\dagger f_l^\dagger f_l \rangle 
	e^{-i (\omega_0 - \omega_k)t} \quad + 2 \sigma^z \sigma^+ \langle f_k^\dagger f_k f_l \rangle 
	e^{i (\omega_0 - \omega_l)t'} \quad + 2 \sigma^z \sigma^- \langle f_k^\dagger f_k f_l^\dagger \rangle 
	e^{-i (\omega_0 - \omega_l)t'} \quad + 4 \sigma^z \sigma^z \langle f_k^\dagger f_k f_l^\dagger f_l \rangle \; \Big] \,
	\rho^S_{\mathrm{int}}(t). 
	\label{C2}
\end{align}

		{
			
			Next, we evaluate the bath correlation function \( \overline{J_k(z) J_k(t')} \) using Wick’s theorem. We consider both bosonic and fermion forms of the environment, though our model ultimately assumes a fermion bath in the rotated basis.
			
			\subsection*{ Wick’s Theorem: General Statement}
			
			Wick’s theorem is a powerful method to compute the expectation values of time-ordered products of creation and annihilation operators in Gaussian states (including vacuum, thermal, and coherent states). It allows such products to be reduced to sums over all possible contractions (i.e., two-point correlation functions)\cite{70}.Let \( A_1, A_2, \dots, A_n \) be bosonic or fermion operators with vanishing odd moments in the Gaussian state. Then:			
			
			(a)For even \( n = 2m \), the expectation value is given by
			\begin{eqnarray*}
			\langle A_1 A_2 \cdots A_{2m} \rangle = \sum_{\text{all pairings}} \text{sign}(\pi) \prod_{(i,j)} \langle A_i A_j \rangle
			\end{eqnarray*}
			(b) For odd \( n \), the expectation vanishes:
			\begin{eqnarray*}
			\langle A_1 A_2 \cdots A_{2m+1} \rangle = 0
			\end{eqnarray*}
			
			Here, \( \text{sign}(\pi) \) accounts for operator permutations (only for fermions), and the sum is taken over all possible ways of pairing the operators into disjoint sets.
						For bosonic operators, contractions commute, while for fermions, anti-commutation relations must be considered.
						In our model, the environment is described by fermion annihilation and creation operators \( f_k \), and we assume that the environment is initially in the vacuum state:
			\begin{eqnarray*}
			|0_B\rangle =\prod_{k=1}^N |0\rangle_k \quad \Rightarrow \quad \langle f_k^\dagger f_k \rangle_B = n_k = 0
			\end{eqnarray*}
						Odd-order moments vanish automatically due to Gaussian nature:
			\begin{eqnarray*}
			\langle f^\dagger f f \rangle_B = 0
		\end{eqnarray*}
			since
			\begin{eqnarray*}
			\langle f^\dagger f f \rangle_B = \langle f^\dagger f \rangle_B \langle f \rangle_B + \langle f^\dagger \rangle_B \langle f f \rangle_B = 0
			\end{eqnarray*}
			as \( \langle f \rangle = 0 \) and \( f^2 = 0 \) (fermion anti-commutation). 			
			Next, we compute the four-point correlation function:
			\begin{eqnarray*}
				\langle f_k^\dagger f_k f_l^\dagger f_l \rangle_B &=& \langle f_k^\dagger f_k \rangle_B \langle f_l^\dagger f_l \rangle_B - \langle f_k^\dagger f_l^\dagger \rangle_B \langle f_k f_l \rangle_B + \langle f_k^\dagger f_l \rangle_B \langle f_k f_l^\dagger \rangle_B \\
				&=& n_k n_l - \langle f_k^\dagger f_l^\dagger \rangle_B \langle f_k f_l \rangle_B + \delta_{kl} n_k (1 - n_k)
			\end{eqnarray*}			
			Given \( n_k = 0 \) in the vacuum state, we find:
			\begin{eqnarray*}
			\langle f_k^\dagger f_k f_l^\dagger f_l \rangle_B = 0
			\end{eqnarray*}
						Therefore, all higher-order terms contributing to the ensemble average \( \overline{J_k(z) J_k(t')} \), which involve such multi-operator correlations, vanish identically. 
			}
			
Therefore, equation \ref{C2} becomes \begin{align}
	& \mathrm{Tr}_B \big\{ \mathbf{H}_{\mathrm{int}}(t) \, \mathbf{H}_{\mathrm{int}}(t') \, 
	\rho^S_{\mathrm{int}}(t) \otimes |0_B\rangle\langle 0_B| \big\}= \frac{1}{N} \sum_{k,l} J_k(t) J_l(t') \Big[ 
	\sigma^+ \sigma^+ \langle f_k f_l \rangle \, e^{i (\omega_0 - \omega_k)(t + t')} \nonumber \\
	&\quad + \sigma^+ \sigma^- (1 - n_k) \delta_{kl} \, e^{i (\omega_0 - \omega_k)(t - t')} \quad + \sigma^- \sigma^+ n_k \delta_{kl} \, e^{-i (\omega_0 - \omega_k)(t - t')} \quad + \sigma^- \sigma^- \langle f_k^\dagger f_l^\dagger \rangle \, e^{-i (\omega_0 - \omega_k)(t + t')} \Big] 
	\rho^S_{\mathrm{int}}(t) \nonumber \\
	&= \frac{1}{N} \sum_{k} J_k(t) J_k(t') \, \sigma^+ \sigma^- \, 
	\rho^S_{\mathrm{int}}(t) \, e^{i (\omega_0 - \omega_k)(t - t')} \label{eq:reduced_corr}
\end{align}

	Similarly, we obtain all other elements and can, therefore write the master equation in the Schrodinger picture as:
	\begin{eqnarray*}
		\dot{\rho}^S(t)= -i [\mathbf{h}_S(t),\rho^S] + \gamma(t)[2\sigma^- \rho^S \sigma^+ - \{\sigma^+ \sigma^-,  \rho^S\}]
	\end{eqnarray*}
	
sd	\section{Deriving solution given in Eq. (\ref{dm})  for the Lindblad master equation in Eq. (\ref{MAS1}) of the main text}\label{apn:Sol}
	
	From Eq. (\ref{MAS1}), we can calculate the elements of evolved density matrix of system as follows:
	{
\begin{align*}
	\langle \uparrow |\frac{\partial }{\partial t} \rho^S(t) |\uparrow\rangle &= \langle \uparrow| -i [\mathbf{h}_S(t), \rho^S(t)] \, |\uparrow\rangle \quad +  \langle \uparrow|\gamma(t)\Big[ 2\sigma^{-} \rho^S(t) \sigma^{+} - \sigma^{+} \sigma^{-} \rho^S(t)  - \rho^S(t) \sigma^{+} \sigma^{-} \Big] |\uparrow\rangle
\end{align*}

	Using the properties of Pauli matrices, the first term can be simplified as
	{\small
		\begin{align*}
			& \langle \uparrow | -i [\mathbf{h}_S(t), \rho^S(t)] | \uparrow \rangle =  \quad \langle \uparrow | \Big( \omega_r \sigma^z - \left[ \int_\uparrow^{t} dt' \sum_k J_k(t) J_k(t') \sin \Delta_k(t - t') \right] \sigma^- \sigma^+ \Big) \rho^S(t) | \uparrow \rangle \nonumber \\
			& \quad - \langle \uparrow | \rho^S(t) \Big( \omega_r \sigma^z - \left[ \int_\uparrow^{t} dt' \sum_k J_k(t) J_k(t') \sin \Delta_k(t - t') \right] \sigma^- \sigma^+ \Big) | \uparrow \rangle  \\
		&=	~~0
		\end{align*}
	}
	
	Similarly,
	\begin{eqnarray*}
		& \langle \uparrow | 2 \gamma(t)\sigma^{-} \rho^S(t) \sigma^{+}|\uparrow\rangle = 0,~~
		\langle \uparrow |  \sigma^{+}\sigma^{-} \rho^S(t)|\uparrow\rangle =  \langle \uparrow | \rho^S(t)|\uparrow\rangle ,~~ 
		\langle \uparrow |  \rho^S(t) \sigma^{+}\sigma^{-} |\uparrow\rangle =  \langle \uparrow | \rho^S(t)|\uparrow\rangle. 
	\end{eqnarray*}
	Therefore, we can write 
	\begin{eqnarray*}
		\frac{\partial }{\partial t} \rho_{\uparrow\uparrow}^S(t) = - 2 \gamma(t) \rho_{\uparrow\uparrow}^S(t), \\ \implies  \rho_{\uparrow\uparrow}^S(t) = e^{-2 \int_0^{t} \gamma(s) ds  } \rho_{\uparrow\uparrow}^S(0) = |a|^2 F_0(t) .
	\end{eqnarray*}}
	Following a similar approach, the remaining elements of the density matrix can also be computed.
	\section{Explicit forms of \texorpdfstring{\(\Gamma(t)\)}{Gamma(t)} and \texorpdfstring{\(\Phi(t)\)}{Phi(t)} for the models considered in the main text}
	\label{apn:GammaPhi}
	\begin{itemize} 
		\item \textit{Exponential Memory Kernel}:\\
		\\
		 Let $\overline{J_k(t) J_k(t')}= \kappa e^{-\kappa(t-t')} $ and by defining the following parameters $ \omega t = \tau , \Bbbk= \kappa/\omega$ and $ \bar{\Delta}= \Delta/\omega$ we calculate
	\begin{eqnarray}
		\Gamma(\tau) &=& \frac{\Bbbk^2}{\Bbbk^2 + \bar{\Delta}^2} \tau 
		- \frac{\Bbbk^3 - \Bbbk \bar{\Delta}^2}{(\Bbbk^2 + \bar{\Delta}^2)^2} \Big[ 1 - e^{-\Bbbk \tau} \cos \bar{\Delta} \tau \Big] \nonumber \\
	\Phi(\tau) &=& \frac{\Bbbk \bar{\Delta}}{\Bbbk^2 + \bar{\Delta}^2} \tau 
	- \frac{\Bbbk^2}{(\Bbbk^2 + \bar{\Delta}^2)^2} \Bigg[ 2 \bar{\Delta} - e^{-\Bbbk \tau} \Big( \frac{\Bbbk^2 - \bar{\Delta}^2}{\Bbbk} \sin \bar{\Delta} \tau + 2 \bar{\Delta} \cos \bar{\Delta} \tau \Big) \Bigg]
	\end{eqnarray}
		\item \textit{Modulated Decay}:\\ 
	
		Let  $\langle J_k(t) J_k(t') \rangle= \kappa e^{-\kappa(t-t')} \cos \omega(t-t') $. In this case we have
	\begin{eqnarray}
		\Gamma(\tau) &=& \frac{\Bbbk}{2}\biggl\{ 
		-\frac{\Bbbk^2 - (\bar{\Delta} + 1)^2}{(\Bbbk^2 + (\bar{\Delta} + 1)^2)^2} 
		- \frac{\Bbbk^2 - (\bar{\Delta} - 1)^2}{(\Bbbk^2 + (\bar{\Delta} - 1)^2)^2} \quad + \Big( \frac{1}{\Bbbk^2 + (\bar{\Delta} + 1)^2} 
		+ \frac{1}{\Bbbk^2 + (\bar{\Delta} - 1)^2} \Big) \Bbbk \tau \nonumber \\
		&& + e^{-\Bbbk \tau} \Bigg[ \frac{\Bbbk^2 - (\bar{\Delta} + 1)^2}{(\Bbbk^2 + (\bar{\Delta} + 1)^2)^2} \cos{(\bar{\Delta} + 1) \tau}\quad - \frac{2 \Bbbk (\bar{\Delta} + 1)}{(\Bbbk^2 + (\bar{\Delta} + 1)^2)^2} \sin{(\bar{\Delta} + 1) \tau}  + \frac{\Bbbk^2 - (\bar{\Delta} - 1)^2}{(\Bbbk^2 + (\bar{\Delta} - 1)^2)^2} \cos{(\bar{\Delta} - 1) \tau} \nonumber \\
		&&- \frac{2 \Bbbk (\bar{\Delta} - 1)}{(\Bbbk^2 + (\bar{\Delta} - 1)^2)^2} \sin{(\bar{\Delta} - 1) \tau} \Bigg] \biggr\}
\\
		\Phi(\tau) &=& \frac{\Bbbk}{2}\biggl\{ \Bigg( 
		\frac{\bar{\Delta}+1}{\Bbbk^2 + (\bar{\Delta}+1)^2} 
		+ \frac{\bar{\Delta}- 1}{\Bbbk^2 + (\bar{\Delta}-1)^2}
		\Bigg) \tau \quad + e^{-\Bbbk \tau} \Bigg[ 
		\frac{2 \Bbbk (\bar{\Delta}+1)}{(\Bbbk^2 + (\bar{\Delta}+1)^2)^2} \cos{(\bar{\Delta}+1) \tau} \nonumber \\
		&& + \frac{\Bbbk^2 - (\bar{\Delta}+1)^2}{(\Bbbk^2 + (\bar{\Delta}+1)^2)^2} \sin{(\bar{\Delta}+1) \tau} \quad + \frac{\Bbbk^2 - (\bar{\Delta}-1)^2}{(\Bbbk^2 + (\bar{\Delta}-1)^2)^2} \sin{(\bar{\Delta}- 1) \tau}  + \frac{2 \Bbbk (\bar{\Delta}-1)}{(\Bbbk^2 + (\bar{\Delta}-1)^2)^2} \cos{(\bar{\Delta}- 1) \tau}
		\Bigg] \nonumber \\
		&& - \frac{2 \Bbbk (\bar{\Delta}+1)}{(\Bbbk^2 + (\bar{\Delta}+1)^2)^2} 
		- \frac{2 \Bbbk (\bar{\Delta}-1)}{(\Bbbk^2 + (\bar{\Delta}-1)^2)^2} 
		\biggr\}~~~~
	\end{eqnarray}
	
	\end{itemize}

\end{document}